\documentclass[usenatbib,usegraphicx,twocolumn]{mn2e}
\usepackage[latin1]{inputenc}
\usepackage{amsmath}
\usepackage{amssymb}
\usepackage{natbib}
\usepackage[]{graphicx}
\usepackage{url}
\title[Cosmic Variance Limited BAO]{Cosmic variance limited Baryon Acoustic Oscillations from the DEUS-FUR $\Lambda$CDM simulation}

\author[Y. Rasera,  P.-S. Corasaniti, J.-M. Alimi, V. Bouillot, V. Reverdy, I. Balmes]{Y. Rasera\thanks{email:yann.rasera@obspm.fr}, P.-S. Corasaniti, J.-M. Alimi, V. Bouillot, V. Reverdy, and I. Balm\`es\\
CNRS, Laboratoire Univers et Th\'eories (LUTh), UMR 8102 CNRS, Observatoire de Paris, \\
Universit\'e Paris Diderot- Paris 7 ; 5 Place Jules Janssen, 92190 Meudon, France}

\begin{document}
\date{}

\maketitle
\label{firstpage}

\begin{abstract} 

We investigate the non-linear evolution of Baryon Acoustic Oscillations (BAO) in the low-redshift matter power spectrum from the DEUS-FUR $\Lambda$CDM model simulation. This is the first cosmological N-body simulation encompassing the full observable cosmic volume, thus allowing cosmic variance limited predictions at BAO scales. We control the effect of numerical systematic errors using a series of large volume high-resolution simulations. The combined analysis allows us to measure the matter power spectrum between $z=0$ and $1$ to $1\%$ over the entire BAO range, $0.03<k~[\textrm{h~Mpc}^{-1}]<0.3$, in bins of size $\Delta k/k\lesssim 1\%$. We define the BAO with respect to a non-linearly evolved wiggle-free spectrum and determine the characteristics of the BAO without recurring to extrapolation from global fitting functions. We quantify the effects of non-linearities on the position and amplitude of the BAO extrema, and the coupling to the broadband slope of the power spectrum. We use these estimates to test non-linear predictions from semi-analytical models. Quite remarkably from the analysis of the redshift evolution of BAO we find that the second dip and third peak remains unaltered by non-linear effects. Furthermore, we find that the square of the damping factor and the shift of the position of BAO extrema scale to good approximation as the square of the growth factor, in agreement with expectations from perturbation theory. This confirms the idea that, besides cosmic distances, an accurate measurement of BAO at different redshifts can directly probe the growth of cosmic structures.

\end{abstract}
  
\begin{keywords}  
Cosmology; N-body Simulations; Large-Scale Structures;
\end{keywords}

\section{Introduction}\label{intro}
The propagation of primeval acoustic waves in the coupled photon-baryon plasma before recombination \citep{sakharov,silk,peeblesyu} generates a distinct pattern of temperature and polarization anisotropies in the Cosmic Microwave Background (CMB). These have been measured with unprecedented accuracy by observations of the Wilkinson Microwave Anisotropy Probe (WMAP) \citep{spergel03,spergel07,komatsu09} and more recently by the Planck satellite \citep{planck13}. 

A similar imprint is present in the late time distribution of large-scale structures in the form of an oscillatory pattern in the matter power spectrum, the so-called Baryon Acoustic Oscillations (BAO). These have been detected using measurements of two-point galaxy correlation function from the Sloan Digital Sky Survey (SDSS) \citep{eisenstein05,huetsi06} and the galaxy power spectrum from the 2-degree Field (2dF) survey \citep{cole05}. 

CMB observations have accurately determined the distance travelled by acoustic waves at decoupling (i.e. the sound horizon). Thus, the detection of BAO in the galaxy distribution can be used as a standard ruler to estimate the angular diameter distance as well as the Hubble rate at redshifts probed by galaxy surveys \citep[see e.g.][]{eisensteinhutegmark98,blake03,seo03}. A new generation of galaxy surveys has been designed to measure the BAO at different redshifts to few percent error from which it will be possible to infer stringent bounds on the cosmological parameters \citep[see e.g.][]{lsst08,euclid2012,boss2013}. This demands for equally accurate cosmological model predictions.

In the linear regime of cosmic structure formation BAO appear as a series of damped oscillations superimposed to the broadband shape of the matter power spectrum. These can be computed to the desired level of accuracy using linear perturbation theory \citep{eisenstein98,lewis00}. However, at low redshifts the onset of the non-linear clustering of matter induces deviations from the linear theory which are much harder to predict. Such non-linearities degrade the BAO pattern by altering the amplitude, shape and position of the extrema as well as the broadband slope of the power spectrum. If not accounted these effects can alter the cosmological parameter inference \citep[see e.g.][]{angulo08}. 

Non-linear effects have been investigated using numerical N-body simulations in numerous studies \citep[e.g.][]{seo05,seo08,angulo08,jeong09,carlson09,nishimichi09,seo10,orban11}. Alternatively, semi-analytical approaches have been developed to compute the non-linear modifications of the matter power spectrum at intermediate scales \citep{nishimichi07,crocce08,padmanabhan09,sherwin12,taruya12}. The comparison with N-body simulations has shown these to provide accurate predictions on BAO scales below $k< 0.1$~h$^{-1}$Mpc \citep[see e.g.][]{carlson09,nishimichi09}. 

However, BAO occur on a relatively large scale ($\sim 100$~h$^{-1}$Mpc) with a shape that depends on small-scale lagrangian displacements $\sim 1-10$~h$^{-1}$Mpc \citep{eisenstein07}. Hence, the main limitation of numerical studies is due to sample variance errors on the one hand and the lack of large dynamical range on the other hand. For instance, \citet{seo10} have used a series of 5000 simulations with $n_{\textrm{part}}=256^3$ particles with box size $L_{\textrm {box}}=1$~h$^{-1}$Gpc from \citet{takahashi09} corresponding to an effective volume of $5000$~h$^{-1}$Gpc$^3$. This allows them to drastically reduce sample variance errors on the matter power spectrum, nevertheless these are still nearly twice as large as cosmic variance uncertainty, while BAO remain under sampled in Fourier space. Because of this, N-body studies of the effects of non-linearities on the BAO have been mainly estimated from functional fits to the numerical power spectrum. Furthermore, averaging over a large ensemble of simulations does not resolve the problem of numerical systematics that needs to be carefully addressed especially when statistical errors are reduced to the percent level. These can be particularly important in the case of low mass and spatial resolution runs such as those from \citet{takahashi09}. To date, N-body simulation studies of the BAO accurate to per-cent level are still missing.

The work presented here aims to fill this gap. To this purpose we use a large volume N-body simulation of a flat $\Lambda$CDM cosmology from the Dark Energy Universe Simulation - Full Universe Runs (DEUS-FUR) \citep{alimi12}. This is one of the three DEUS-FUR simulations of Dark Energy models with $8192^3$ particles and box length of $21$~h$^{-1}$Gpc. The size of the simulation box allows us to derive cosmic variance limited predictions for the matter power spectrum. However, in order to control the effect of numerical systematics we perform a convergence study using a set of N-body simulations for which we have varied several numerical simulation parameters. 

The combined analysis allows us for the first time to determine the effect of non-linearities on all BAO peaks and troughs between redshift $z=0$ and $z=1$ to $1\%$ accuracy level (and in bins of size $\Delta k/k\lesssim 1\%$). It is in this redshift interval that the power spectrum is most difficult to predict. On the other hand it is the probed range of current and future observational missions (to give an idea the median redshift of the planned Euclid survey is $\approx 1$). 

We use the DEUS-FUR power spectrum to test the validity of semi-analytical models on the BAO scale. In addition, we precisely characterize the non-linear modifications of the BAO extrema without resorting to curve-fitting procedures. We show that the shift of the position of the BAO extrema remains a few percent effect at all redshift. In contrast, the damping of the BAO amplitude is the dominant modification induced by the non-linear clustering of matter. This evolves as function of redshift proportionally to the linear growth rate, thus suggesting that measurements of the amplitude of BAO extrema at different redshifts can be used as an independent cosmological probe carrying information on the linear growth of structures. 

The paper is organized as follows. In Section~\ref{nbody} we summarize the characteristics of the N-body simulations used in our analysis, while in Section~\ref{error} we present a detailed study of the statistical and systematic errors affecting the measurement of the matter power spectrum. In Section~\ref{BAO} we define the BAO spectrum and discuss the comparison against semi-analytical models. In Section~\ref{BAO_NL} we quantify the non-linear effects on the position and amplitude of the BAO extrema and the coupling to the broadband slope of the power spectrum. Finally, we discuss our conclusion in Section~\ref{conclu}.

%%%%%%%%%%%%%%%%%%%%% SECTION 2 %%%%%%%%%%%%%%%%%%%%%%%%%%%%%%%%%%%%%%%%%%%%%%%%

\section{N-body simulations}\label{nbody}
\subsection{Simulation set}
We use the DEUS-FUR simulation of a flat $\Lambda$CDM cosmology best-fit to the WMAP-7yr data \citep{spergel07}. This has been realized using the application AMADEUS -- A Multi-purpose Application for Dark Energy Universe Simulation -- expressly developed for the DEUS-FUR project \citep{alimi12}. Gaussian initial conditions using the Zel'dovich approximation have been generated with an optimized version of the code MPGRAFIC \citep{prunet08}. The N-body run has been performed with a version of RAMSES \citep{teyssier02} that has been specifically improved to run on a large number of cores ($\ge 40000$). This is an Adaptive Mesh Refinement (AMR) code with a tree-based structure that allows for recursive refinements above a user-defined density threshold. The N-body solver evolves particles using a Particle-Mesh (PM) method, while the Poisson equation is solved with a multigrid technique \citep{guillet11}. A detailed description of the algorithms, optimization schemes and the computing challenges involved with the realization of DEUS-FUR is given in \citet{alimi12}, while the characteristics of the three DEUS-FUR simulations of Dark Energy models will be presented in a forthcoming paper. 

The box length of the DEUS-FUR simulations is set to $21\,$h$^{-1}$Gpc, thus enclosing the horizon diameter of the three simulated cosmologies ($d_H\approx 20.7\,$h$^{-1}$Gpc in $\Lambda$CDM). 

These simulations employ $8192^3$ particles and a coarse grid of $8192^3$ resolution elements with $6$ levels of refinement, reaching a formal mass resolution of $1.2 \times 10^{12}$~h$^{-1}$M$_\odot$ and a spatial resolution of $40\,$h$^{-1}$kpc (corresponding roughly to the mass and size of the Milky Way). 

The initial redshift is set to $z_i=106$, hence sufficiently high to avoid transient effects which occur in the case of a late start of the initial conditions using the Zel'dovich approximation \citep{scoccimarro98,crocce06}. The refinement threshold is set to $m_\textrm{ref}=14$ such as to limit the total amount of AMR cells generated during the run (and the associated memory usage) without affecting the accuracy of the non-linear power spectrum calculation. The final count of AMR cells is still of $\sim 2$ trillions. The model parameters of the simulated $\Lambda$CDM-W7 are quoted in Table~\ref{cosmo}, while the characteristics of the DEUS-FUR simulation are listed in Table~\ref{deusfur}.

\begin{table} 
\begin{center}
\begin{tabular}{cccccc}
\hline\hline
Model &$\Omega_{m}$ & h & $\sigma_8$ & $n_s$ & $\Omega_b$ \\
\hline
$\Lambda$CDM-W7&0.2573&0.72 & 0.801&0.963&0.04356\\
\hline
\end{tabular}
\caption{DEUS-FUR $\Lambda$CDM-W7 model parameter values. \label{cosmo}}  
\end{center}
\end{table}

\begin{table*} 
\begin{center}
\begin{tabular}{ccccccccccc}
\hline\hline
L$_{\textrm{box}}$ (h$^{-1}$Mpc) &n$_\textrm{part}$ & n$_x$ & n$_{\textrm{ref}}$&m$_\textrm{ref}$&n$_\textrm{cell}$ & m$_p$ (h$^{-1}$M$_\odot$) & $\Delta_x$ (h$^{-1}$kpc)& z$_i$ &C$_{\textrm{dt}}$& P$_\textrm{lin}(k)$  \\
\hline
21000 &$5.5\times10^{11}$&8192& 6 &14&$2\times10^{12}$&$1.2\times10^{12}$&40 &106&0.2&CAMB\\
\hline
\end{tabular}
\caption{\label{deusfur} DEUS-FUR $\Lambda$CDM-W7 model simulation parameters: L$_{\textrm{box}}$ is the box length, n$_{\textrm{part}}$ is the number of particles, n$_x$ the grid size, n$_{\textrm{ref}}$ the number of refinements, m$_\textrm{ref}$ the refinement threshold, n$_{\textrm{cell}}$ the final number of AMR cells, m$_p$ the particle mass, $\Delta_x$ the spatial resolution, z$_i$ the starting redshift, C$_{dt}$ is the Courant-like factor which determines the size of the integration time-step (see text for details) and P$_\textrm{lin}(k)$ specifies the source of the linear power spectrum used to generate initial conditions. All calculations have been performed in double precision.}  
\end{center}
\end{table*}

\begin{table} 
\begin{center}
\begin{tabular}{ccccc}
\hline \hline
L$_{\textrm{box}}$  & n$_x$ & m$_{\textrm{ref}}$  & z$_i$ &C$_{\textrm{dt}}$  \\
(h$^{-1}$Mpc)          &       &                 &       &   \\  
\hline
$10500$ &$4096$&$14$&$106$&0.2\\
$5250$  &$2048$&$14$&$106$&0.2\\
$2625$  &$1024$&$14$&$106$&0.2\\
$1312$  &$512$&$14$&$106$&0.2\\
$5250$  &$2048$&$8$&$106$&0.2\\
$5250$  &$2048$&$25$&$106$&0.2\\
$5250$  &$2048$&$14$&$106$&0.08\\
$5250$  &$2048$&$14$&$106$&0.5\\

$5250$  &$2048$& $14$ &$272$ &0.2\\
$5250$  &$2048$& $14$ &$170$ &0.2\\
$5250$  &$2048$& $14$ &$66$ &0.2\\
$5250$  &$2048$& $14$ &$41$ &0.2\\

$2592$  &$2048$& $8$  &$56$  &0.5\\
$2592$  &$1024$& $8$  &$56$  &0.5\\
$648$   &$1024$& $8$  &$93$  &0.5\\
$648$   &$512$ & $8$  &$93$  &0.5\\
$648$   &$256$ & $8$  &$93$  &0.5\\

$2625$  &$1024$& $14$ &$106^{\star}$&0.2\\
$5250$ &$2048$&$14$&$106^{\star \star}$&0.2\\
\hline
\end{tabular}
\caption{\label{deusfurtest} Characteristics of the simulations used for the evaluation of numerical systematic uncertainties. Parameter specifications as in Table~\ref{deusfur}. The initial power spectrum is computed with CAMB and the initial conditions are generated using Zel'dovich approximation. In the case ($^\star$) the initial conditions are generated using second order Lagrangian Perturbation Theory. In the case  ($^{\star \star}$) the initial power spectrum is given by the wiggle-free formula of \citet{eisenstein98}.}  
\end{center}
\end{table}

We generate a set of N-body simulations which we use to perform convergence tests and control the effect of numerical systematics on the DEUS-FUR matter power spectrum. These consist of smaller volume simulations typically with $2048^3$ particles in which we have varied the simulation box length, the refinement threshold, the starting redshift, the generation of initial conditions, the integration time-step and the mass resolution. The characteristics of these simulations are listed in Table~\ref{deusfurtest}. This benchmark is similar to that used in \citet{reed13}, though in our case we have opted for higher-resolution larger-volume simulations. 

In addition to this testbed of simulations, we have performed a N-body run with an initial wiggle-free linear matter power spectrum from \citet{eisenstein98}. We use the resulting spectrum as reference wiggle-free power spectrum to define the BAO. The initial spectra for all other simulations has been computed using the CAMB code \citep{lewis00}. 

\subsection{Power spectrum estimator}
We compute the matter power spectrum from the N-body runs with the code POWERGRID \citep{prunet08} which we have optimized to handle a FFT grid with $\sim 4$ trillion elements. The FFT grid size is two times thinner than the coarse grid of the dynamical solver. The resulting power spectrum is corrected for the effect of smoothing due to the Cell-In-Cell (CIC) method. We do not perform any correction for the shot noise, which turns out to be completely negligible. Similarly, we do not correct for aliasing, since, by varying the size of the CIC grid we find that aliasing effects are negligible below half the Nyquist frequency of the CIC grid. This sets the computation of the power spectrum in the interval $k_{\textrm{min}}=2 \pi /L_{\textrm{box}}$ and $k_{\textrm{max}}=\pi/ \Delta^{\textrm{coarse}}_x$, where $L_{\textrm{box}}$ is the simulation box length and $\Delta^{\textrm{coarse}}_x$ is the spatial resolution at the coarse grid level. 

\begin{figure*}
\begin{center}
\includegraphics[width=0.8\hsize]{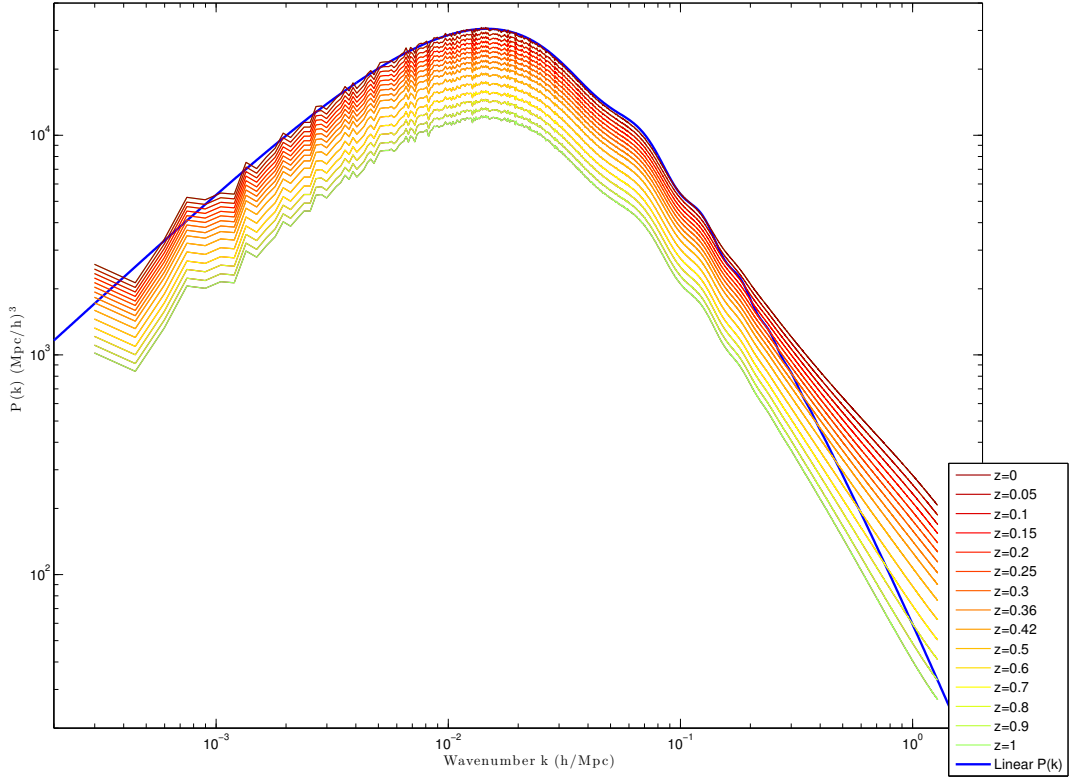}
\end{center}
\caption{DEUS-FUR $\Lambda$CDM-W7 real-space matter power spectrum at 15 redshifts between redshift $z=0$ and $1$. The wavenumber interval almost spans four decades, 
from $k_{\textrm{min}}=2 \pi/L_{\textrm{box}}\approx 2\pi/d_{\textrm{H}}$ (where $L_{\textrm{box}}$ is the box length and $d_{\textrm{H}}$ is the diameter of the observable universe) to the Nyquist frequency of the coarse grid. The solid blue line corresponds to the linear matter power spectrum at $z=0$.}  
\label{powerallz}
\end{figure*}

%%%%%%%%%%%%%%%%%%%%% SECTION 3 %%%%%%%%%%%%%%%%%%%%%%%%%%%%%%%%%%%%%%%%%%%%%%%%

\section{Matter power spectra: statistical and systematic errors}\label{error}
We now present the DEUS-FUR $\Lambda$CDM-W7 power spectrum and discuss the evaluation of statistical and systematic errors. 

Figure~\ref{powerallz} illustrates the evolution of the matter power spectrum between redshift $z=0$ to $1$ over a range of scales spanning nearly four order of magnitudes, from $k_{\textrm{min}}=3 \times 10^{-4}$~h Mpc$^{-1}$ to  $k_{\textrm{max}}=1.2$ h Mpc$^{-1}$. 

One remarkable feature of the plotted spectra is the low level of noise. Because of the large size of the simulation box the noise is $<1\%$ even on scales near the peak of the CDM power spectrum at $k\simeq 10^{-2}$~h Mpc$^{-1}$. We may also notice the excess of power at small scales ($k>10^{-1}$~h Mpc$^{-1}$) due to the onset of the non-linear clustering regime. At $z=0$ this alters by $\sim 1\%$ the prediction of the linear theory up to very large scales $k\simeq 10^{-2}$~h Mpc$^{-1}$. The imprint of the BAO is clearly distinguishable in all plotted spectra, extending over a decade from $k= 3\times 10^{-2}$~h$^{-1}$Mpc to $k= 3\times 10^{-1}$~h$^{-1}$Mpc with a peak-to-peak amplitude of $\sim 10\%$. 

Numerical effects on the matter power spectrum have already been investigated in several works \citep[see e.g.][]{oshea05,heitmann05,heitmann08,joyce09,heitmann10,smith12,reed13}. Such studies have shown that the nature and amplitude of numerical systematics may depend on the specifics of the N-body solver. Hence, it is critical for us to investigate numerical source of errors which may be specific to the solver we have used. 

In principle, we could simply focus on the power spectrum normalized by the spectrum of the initial conditions. This would allow us to suppress the amplitude of statistical fluctuations on the large scale, but it will not reduce sample variance errors nor discount for possible numerical systematics, especially at small scales where non-linearities induce phase correlations in Fourier space.

The error analysis that we present here extends previous studies to the case of very large volume simulations ($>5~$h$^{-1}$Gpc) with large number of particles ($\sim 2048^3$). Our goal is to evaluate the error budget on the DEUS-FUR matter power spectrum and select the range of scales where the power spectrum is determined to $<1\%$ accuracy. 

\subsection{Statistical errors}
The finite size of cosmological simulations introduces two types of statistical errors. Firstly, it reduces the number of accessible modes causing sample variance errors which dominate the error budget when probing scales near the size of the simulation box. Secondly, modes which are larger than the simulation box length are absent \citep{bagla05} and since the gravitational coupling to these modes is missing, this results in a lower amplitude of the power spectrum \citep{heitmann10}. Both problems can be handled either by averaging the spectra of a large number of different realizations \citep[e.g.][]{takahashi08,takahashi09} or setting the simulation box length to be as large as possible. However, since observations are limited by the size of the cosmological horizon, statistical errors cannot be reduced to less than cosmic variance. Hence, by setting the box length of the DEUS-FUR simulations to the diameter of the observable universe we are guaranteed to derive cosmic variance limited (i.e. minimal sample variance) predictions for the matter power spectrum. 

\begin{figure}
\includegraphics[width=0.9\hsize]{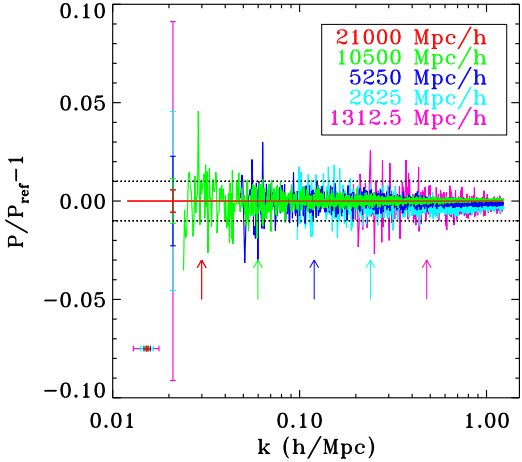}
\caption{Relative difference of the matter power spectra at $z=0$ from testbed simulations of different box lengths to that from DEUS-FUR $\Lambda$CDM-W7 with $L_{\textrm{box}}=21000~$h$^{-1}$Mpc (red line). The various lines corresponds to the simulation with box length $L_{\textrm{box}}=10500~$h$^{-1}$Mpc (green line), $L_{\textrm{box}}=5250~$h$^{-1}$Mpc (light-blue line), $L_{\textrm{box}}=2625~$h$^{-1}$Mpc (dark-blue line) and $L_{\textrm{box}}=1312.5~$h$^{-1}$Mpc (magenta line) respectively. All curves are truncated outside the range where $\sigma_{\textrm{noise}}/P>0.01$ (see text). The vertical error bars at $k=0.02$~h Mpc$^{-1}$ shows $\sigma_{\textrm{noise}}$, while the horizontal ones give the wavenumber bin size for the different simulation box lengths respectively. The arrows mark the wavenumber values where $dk$ equals to $1\%$ of $k$. }  
\label{pk_box}
\end{figure}

In Figure~\ref{pk_box} we plot the power spectrum at $z=0$ from the auxiliary simulations with $L_{\textrm{box}}$ varying from 1.3~h$^{-1}$Gpc to 10.5~h$^{-1}$Gpc relative to the DEUS-FUR case. As expected, the statistical fluctuations on the power spectrum increase as the simulation box length decreases. 

The root-mean-square fluctuation of the initial power spectrum is given by \citep{jeong09}:
\begin{equation}\label{sigmanoise}
\sigma_{\textrm{noise}}(k)=\sqrt{\frac{2}{N_{\textrm{modes}}}}\left[P(k)+\frac{1}{N_{\textrm{mean}}}\right],
\end{equation}
where $N_{\textrm{modes}}=4 \pi k^2 /dk^2$ is the number of modes in a shell of size $dk=2 \pi/L_{\textrm{box}}$ and $N_{\textrm{mean}}$ is the mean particle number density. The propagation of this error into the non-linear regime is non-trivial \citep{ngan12}, however we think that Eq.~(\ref{sigmanoise}) still provides a reasonable approximation on the BAO scales. The error bars on the left-hand side of Figure~\ref{pk_box} give a visual comparison of the expected statistical error $\sigma_{\textrm{noise}}$ at $k=0.02$~h Mpc$^{-1}$ for the different simulation box lengths. We can see that even a large volume simulation with box length of $1.3$~h$^{-1}$Gpc still leads to statistical errors of order of $10\%$, whereas for DEUS-FUR these are within $0.5\%$. 

We select the interval of interest by truncating the DEUS-FUR power spectrum over modes where the statistical error given by Eq.~(\ref{sigmanoise}) exceeds the $1\%$ level. This sets a lower bound $k_{\textrm{min}}\approx0.01$~h Mpc$^{-1}$. Notice that for all simulations the measured fluctuations are indeed within $\sim 1\%$ level.

The finite size of the simulation box length is also relevant in determining the accuracy of the mode sampling, which is critical to accurately locate the position of the BAO extrema. In fact, the wavenumber bin-size is given by $dk=2 \pi/L_{\textrm{box}}$, that sets the size of the ``error bars'' along the $k$-axis. These are illustrated in Figure~\ref{pk_box} at $k\approx 0.02$~h Mpc$^{-1}$ for different simulation box lengths. The arrows mark the wavenumber values where $dk$ equals to $1\%$ of $k$ for a given simulation box length. 

As already mentioned, running a large ensemble of small-box simulations \citep[as in the case of][]{takahashi09} is effective to reduce statistical errors, however it does not improve the k-sampling of the spectrum which is (usually) governed by the simulation box length. Common methods to reduce the uncertainty on the wavenumber value and thus the location of the BAO consist in using functional fitting procedures. For instance, one can use a functional form (e.g. polynomial expansions) to best-fit the measured power spectrum and subsequently infer the position of the BAO from the resulting best-fit function. Alternatively, one can bin the power spectrum or convolve it with a filter function. In any case, these procedures do not provide better information than that fixed by the size of $dk$ (i.e. the simulation box length). Therefore the trade-off of these methods is a loss of accuracy in the determination of the location of the BAO peaks and dips, which is key to the determination of cosmological distances. In fact, the position of the BAO as inferred from the best-fitting function may be sensitive to the specific choice of the functional form used to fit the power spectrum. 

In the case of the DEUS-FUR simulations this issue does not arise since the choice of the box length guarantees the ideal $k$-sampling. 

\subsection{Systematic errors}
The determination of numerical systematic effects on the matter power spectrum is critical to reach the $1\%$ accuracy required by future surveys. Here, we assess their impact by comparing the DEUS-FUR power spectrum with that obtained from the testbed simulations in which we have varied the mass resolution, the refinement strategy, the starting redshift, the generation of the initial conditions and the integration time-step. The results of this comparison are shown in Figure~\ref{systematic}.

\subsubsection{Mass resolution}
The mass resolution is a key attribute of N-body simulations. To date only a few studies have investigated the dependence of the matter power spectrum on the mass resolution of N-body simulations \citep[see e.g.][]{joyce09,heitmann10}. A brute force approach consisting in running simulations with fixed volume and increasingly large number of particles is optimal to evaluate the amplitude and mode dependence of this effect. It is in this way that \citet{heitmann10} have found that the power spectrum falls by $\sim 8\%$ at $k=0.3$~h Mpc$^{-1}$ for a mass resolution of $3\times 10^{12}$~h$^{-1}$M$_\odot$. 

\begin{figure*}
\begin{tabular}{cc}
\includegraphics[width=0.45\hsize]{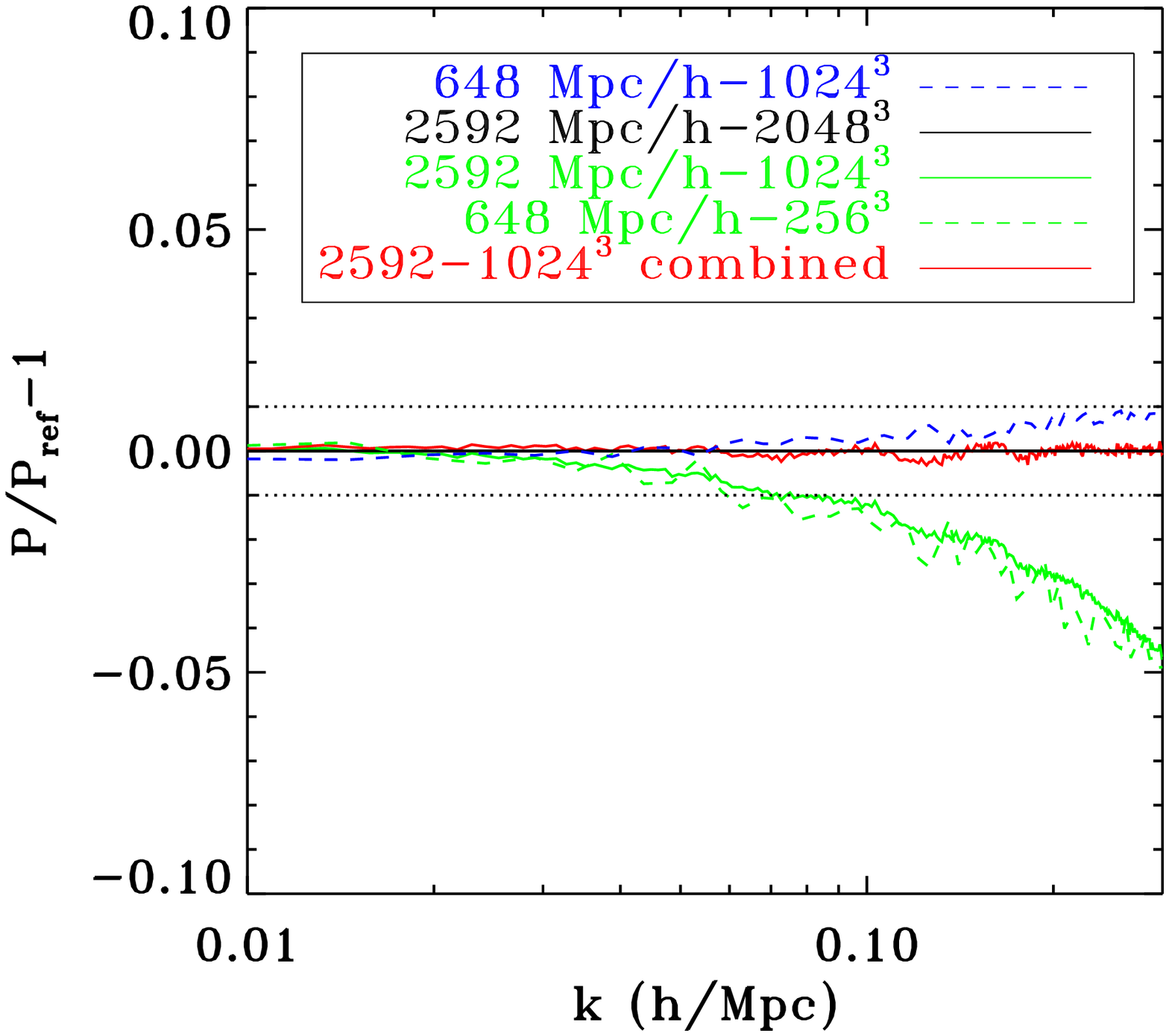}&\includegraphics[width=0.45\hsize]{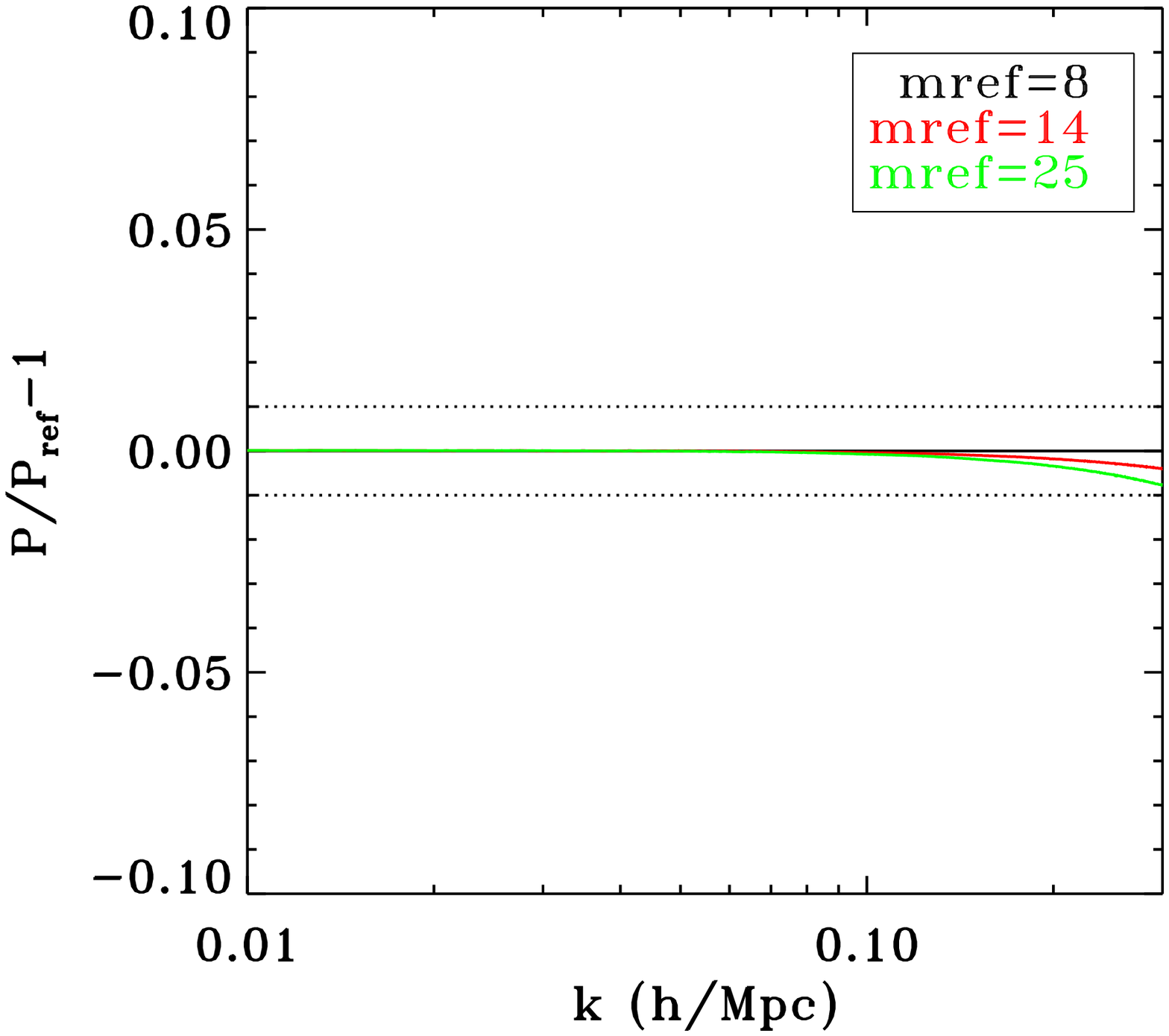}\\
\includegraphics[width=0.45\hsize]{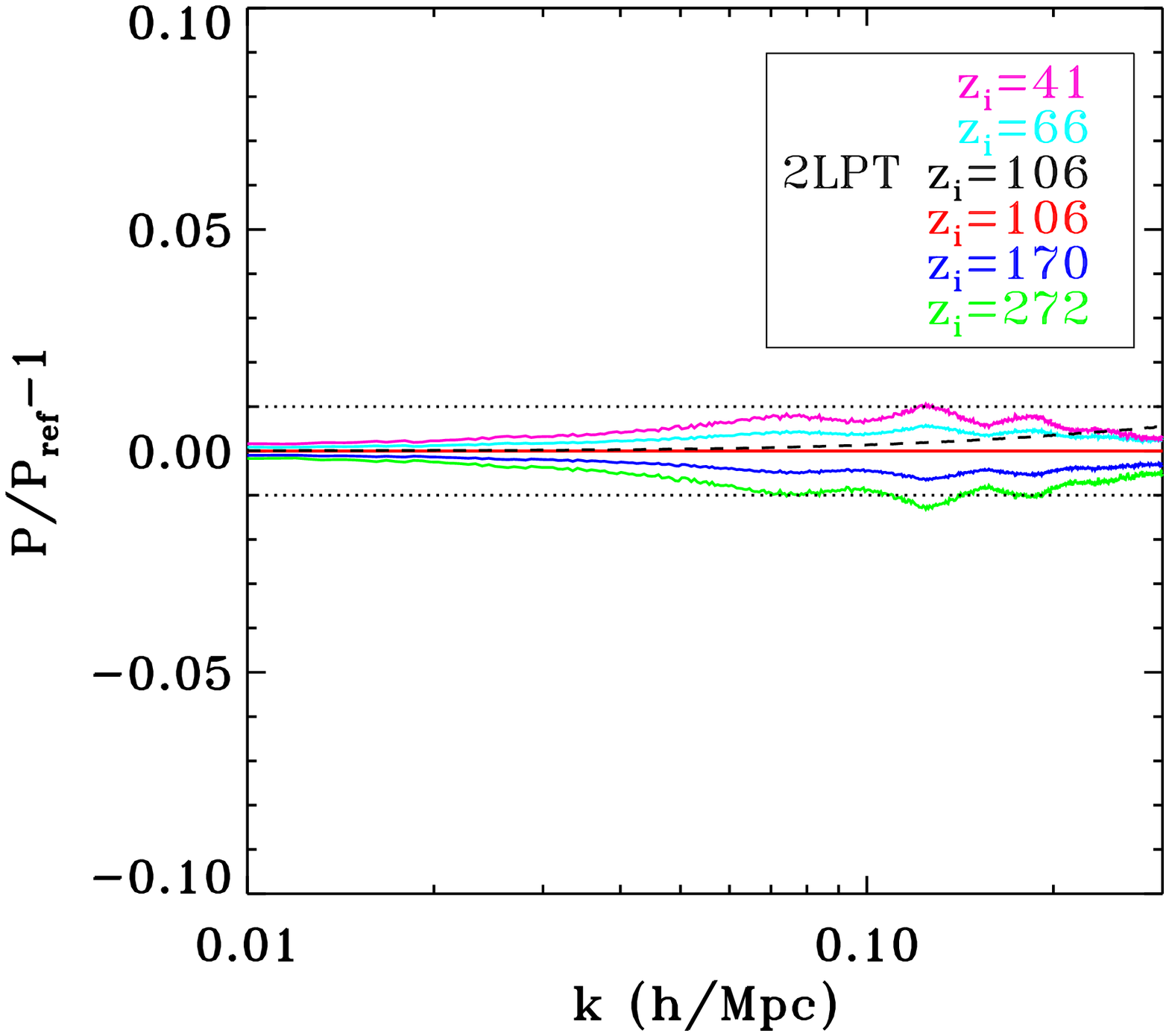}&\includegraphics[width=0.45\hsize]{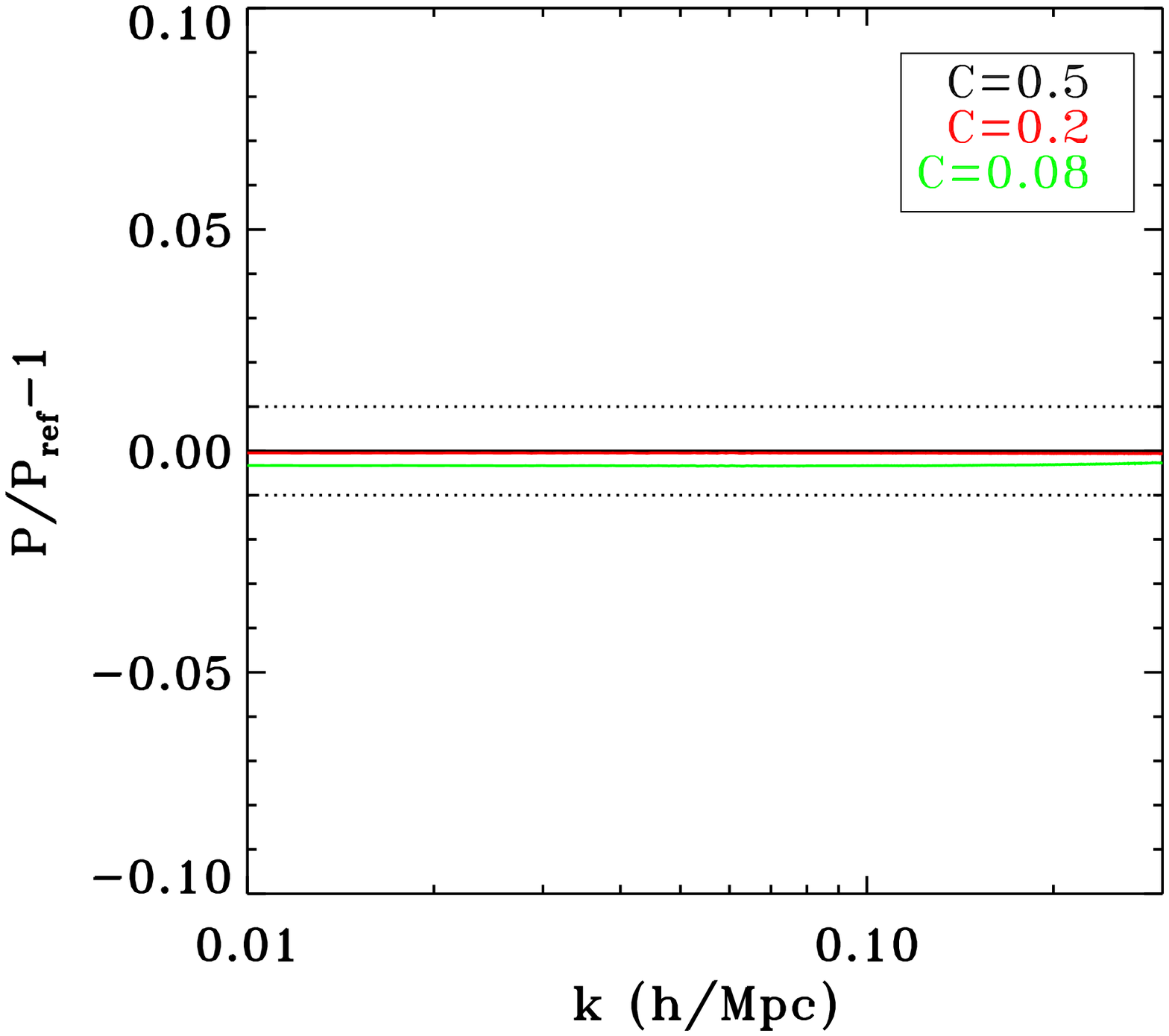}\\
\end{tabular}
\caption{Relative difference of the matter power spectra from the testbed simulations listed in Table~\ref{deusfurtest}. Top-left panel: mass resolution dependence from simulations with $m_p=1.2\times 10^{12}$h$^{-1}$M$_\odot$ (green lines), $m_p=1.5\times 10^{11}$h$^{-1}$M$_\odot$ (black line) and $m_p=1.8\times 10^{10}$h$^{-1}$M$_\odot$ (blue line) respectively. The reference spectrum is given by the simulation with $m_p=1.5\times 10^{11}$h$^{-1}$M$_\odot$. The red line corresponds to spectrum of the simulation with $1024^3$ particles and $2592$~h$^{-1}$Mpc box length corrected for the mass resolution effect estimated by $r_{\textrm{corr}}$ (see text). Top-right panel: dependence on the refinement threshold from simulations with $m_{\textrm{ref}}=8$ (black line), $m_{\textrm{ref}}=14$ as in DEUS-FUR (red line) and $m_{\textrm{ref}}=25$ (green line). Bottom-left panel: dependence on the starting redshift and the generator of initial conditions from simulations with $z_i=272$ (green line), $z_i=170$ (dark-blue line), $z_i=106$ (red line),  $z_i=66$ (blue line), and $z_i=41$ (magenta line) using ZA and $z_i=106$ using 2LPT (black line). Bottom-right panel: dependence on the integration time-step size from simulations with ``large'' (green line), ``medium'' (red line) and ``mall'' (black line). As we can see in all cases the systematic effects on the DEUS-FUR spectrum are within $1\%$ level (red lines) over the range $k=0.01-0.3$~h Mpc$^{-1}$.}
\label{systematic}
\end{figure*}

Here, we perform a similar analysis by comparing the power spectra at $z=0$ from the testbed of different mass resolution simulations. The result of the comparison is shown in the top-left panel of Figure~\ref{systematic}, where we plot the relative difference of the power spectrum (normalized to the linear prediction) of a given simulation relative to that of a reference case with mass resolution $m_p=1.5\times 10^{11}$~h$^{-1}$M$_\odot$ (black solid line) characterized by $L_{\textrm{box}}=2592$~h$^{-1}$Mpc and $2048^3$ particles. The simulations with mass resolution $m_p=1.2\times 10^{12}$~h$^{-1}$M$_\odot$, similar to that of DEUS-FUR have $L_{\textrm{box}}=2592$~h$^{-1}$Mpc with $1024^3$ particles (green solid line) and $L_{\textrm{box}}=648$~h$^{-1}$Mpc with $256^3$ (green dashed line) respectively. We also consider a simulation with a higher-resolution with respect to the reference one with $m_p=1.8\times 10^{10}$~h$^{-1}$M$_\odot$ and characterized by $L_{\textrm{box}}=648$~h$^{-1}$Mpc with $1024^3$ particles (blue dashed line). 

We can see that the lower the mass resolution the larger the deviation of the power spectra at high-k. In particular the reference simulation underestimates the power spectrum with respect to the higher-resolution by $<1\%$ up to $k=0.3$~h Mpc$^{-1}$, while the simulations with the DEUS-FUR resolution underestimates the power spectrum by $5\%$ which is compatible with the results by \citet{heitmann10}. 

This effect is a generic feature of codes used for large N-body simulations which rely on a PM method to solve the large-scale dynamics. The fact that the amplitude of this effect increases with $k$ does not mean that small-scale structures such as halos are not well resolved (some halos have 10000 particles), rather that most of the particles are still in underdense regions where the force calculation is not refined. \citet{knebe01} have performed Zel'dovich wave tests and shown that in void regions $\sim 8$ cells per particle are necessary to accurately follow the Zel'dovich wave, while inside virialized structures at least $\sim 8$ particles per cell are required to suppress the Poisson noise. However, such a refinement strategy is too expensive for large volume simulations such as DEUS-FUR, since it would result in an extremely large number of AMR cells and consequently an excessive memory usage beyond computational capabilities. Nevertheless, one can correct for such an error by combining information from the higher resolution simulations. 

As it can be seen from the plot in the top-left panel of Figure~\ref{systematic} the mass resolution effect on the matter power spectrum is a very smooth function of $k$. It does not alter the BAO structure, but only affects the broadband shape. Hence, we can infer a precise estimate of the mass resolution effect by taking the ratio of the matter power spectrum from the simulation with $2592$~h$^{-1}$Mpc box length and $1024^3$ particles (that has the same mass resolution as DEUS-FUR) to that of the simulation with same box length and $2048^3$ particles (that has a higher resolution with $m_p=1.5\times 10^{11}$~h$^{-1}$M$_\odot$). The correcting function $r_{\textrm{corr}}$ for each redshift is then obtained by fitting this ratio with a polynomial as function of $k$. As we can see in the top-left panel of Figure~\ref{systematic} the power spectrum of the simulation with $1024^3$ particles and $2592$~h$^{-1}$Mpc box length is within $1\%$ of the higher resolution simulation when divided by $r_{\textrm{corr}}$ (red solid line) up to $k_{\textrm{max}}=0.3$~h Mpc$^{-1}$. This sets the upper bound on the wavenumber interval of interest. Thus, the mass resolution effect is corrected in the DEUS-FUR power spectrum by considering  $P^{\textrm{corr}}_{\textrm{DEUS-FUR}}=P_{\textrm{DEUS-FUR}}/r_{\textrm{corr}}$. We have tested that improving the mass resolution by another factor $8$ gives the same spectra at $1\%$ level. Therefore, we are confident that the estimated ratio accurately corrects the mass resolution effect on the DEUS-FUR power spectrum as well.

The end result of this error analysis is the selection of the interval of interest $k=0.01-0.3$~h Mpc$^{-1}$ where the dominant systematic uncertainty due to mass resolution and the statistical errors are both within $1\%$ level. This is an unprecedented achievement that, as we will discuss in the next Section, allows us to precisely evaluate the effects of non-linearities on the BAO.

\subsubsection{Refinement strategy}\label{refeffect}
The AMR algorithm allows for an accurate calculation of the force over a large dynamical range. The AMR-tree dominates the memory usage, since it is nearly proportional to the number of grid cells. Thus, it can be a limiting factor when running large volume high-resolution simulations. This has forced us to optimize the refinement strategy for DEUS-FUR simulations, compromising between the accuracy on the matter power spectrum (and the mass function) and the memory usage. As an example, a simulation with $5.3$~h$^{-1}$Gpc box length and 8.6 billion particles at the end of the run has a total number of cells that varies between $52$ billions for a refinement threshold $m_{\textrm{ref}}=8$ (number of particles per cell) and $18$ billions for a refinement threshold of 25. We have opted for a refinement threshold of 14 (corresponding to a total of 30 billion cells for the testbed simulation). As illustrated in the top-right panel of Figure~\ref{systematic} the effect of the refinement does not alter the matter power spectrum by more than $0.5\%$ over the interval of interest and consequently it can be neglected.

\subsubsection{Initial conditions}\label{iceffect}
The generation of initial conditions and the choice of the starting redshift of the simulations are also potential sources of systematic errors. The effect of transients on the matter power spectrum \citep{scoccimarro98} has been studied in detail by \citet{crocce06}. These authors have shown that starting the initial conditions using the Zel'dovich approximation (ZA) at redshift $z_i<49$ leads to underestimated power spectra at the $3\%$ level (near $k\approx1$~h Mpc$^{-1}$). This can be corrected by generating initial conditions using second order lagrangian perturbation theory (2LPT). On the other hand, \citet{reed13} have shown that using either ZA or 2LPT with a starting redshift $z_i>200$ leads to numerical errors of $20\%$ on the mass function and $15\%$ on the power spectrum (near $k\approx 1$~h Mpc$^{-1}$), whereas a starting redshift between $z_i=30$ and $100$ with 2LPT gives a converged mass function at the few percents level for halos larger than $200$ particles and a converged power spectrum at the percent level. 

In the bottom-left panel of Figure~\ref{systematic} we plot the relative difference of the matter power spectrum for different initial condition generators and starting redshifts. We can see that using ZA or 2LPT at $z_i=106$ makes very little difference ($<0.5\%$). We have also explored a wide range of starting redshifts varying from $41$ to $272$ and found that differences increase only up to $1\%$ level between $k=0.01$ and $0.3$~h Mpc$^{-1}$. Interestingly, for higher starting redshift ($z_i>272$) our PM-AMR N-body solver leads to a lower power spectrum on the large scales ($k<0.3$~h Mpc$^{-1}$). As a result of these tests we have set the DEUS-FUR starting redshift to $z_i=106$.

\subsubsection{Time step}
The integration time-step is another simulation parameter that may affect the accuracy of the matter power spectrum. RAMSES uses an adaptive time-step method where the time-step of the level $\ell +1$ is divided by a factor 2 compared to the time step of the level $\ell$ \citep{teyssier02}. The time-step at the coarse level is given by the minimum between $C_{\textrm dt}H/5$ (where $H$ is the Hubble constant), $C_{\textrm dt}t_{\textrm{ff}}$ (where $t_{\textrm{ff}}$ is the local free-fall time) and $C_{\textrm dt} \Delta_x/v_p^{\textrm{max}}$ (where $\Delta_x$ is the spatial resolution, and $v_p^{\textrm{max}}$ is the maximum velocity of dark matter particles). At low redshift, the latter condition is the more stringent. The default value of the Courant-like factor is $C_{\textrm dt}=0.5$. For comparison we have run testbed simulations with such ``large'' time-step divided by a factor of $2.5$ ($C_{\textrm dt}=0.2$) and $6.3$ ($C_{\textrm dt}=0.08$). In the bottom-right panel of Figure~\ref{systematic} we can clearly see that the effect of these different time-steps on the matter power spectrum remains negligible. In the case of the DEUS-FUR simulations we have set the Courant-like factor to $C_{\textrm dt}=0.2$ since this allows us to increase the time resolution of the lightcone data.

%%%%%%%%%%%%%%%%%%%%% SECTION 4 %%%%%%%%%%%%%%%%%%%%%%%%%%%%%%%%%%%%%%%%%%%%%%%%

\section{Non-Linear Evolution of Baryon Acoustic Oscillations}\label{BAO}

\subsection{BAO Spectrum}
The oscillatory pattern that characterizes the imprint of BAO on the matter power spectrum is usually defined relative to a smooth (wiggle-free) function, $P_{\textrm{smooth}}(k)$, that accounts for the broadband slope of the underlying spectrum. Clearly, the choice of this function becomes critical if we aim to achieve $1\%$ accuracy on the BAO. This is because the relative amplitude of BAO is of order $5-10\%$, hence even a change as small as $1\%$ in the definition of $P_{\textrm{smooth}}(k)$ translates into a $10-20\%$ variation in the BAO amplitude. 

Several approaches have been considered in the literature to define $P_{\textrm{smooth}}(k)$. The simplest choice is to assume a smooth version of the linear power spectrum. For instance \citet{eisenstein98} provides a formula for a wiggle-free power spectrum. This is an unphysical one that nonetheless accounts for the broadband slope of the linear matter power spectrum. However, its use is limited by the fact that the non-linear regime increasingly boosts the amplitude of the spectrum at larger $k$, thus altering the broadband slope of the spectrum with respect to the linear prediction in a way that erases the BAO signature. 

Alternatively, one can define a smooth function directly from observations. For instance, \citet{percival07} use a nine-node cubic spline to fit the broadband slope of the measured power spectrum, having excluded BAO data points from the fit. \citet{seo05} use a similar approach, but instead of a cubic interpolation they assume polynomial fitting functions. Such methods, although free of cosmological assumptions, may introduce spurious effects in the analysis of the BAO. This is because the freedom in the choice of the fitting interval or the specific form of the fitting functions can potentially absorb part of the BAO signal. As we will argue this is inevitable to happen for measurements that aims $1\%$ accuracy due to a subtle coupling between the BAO and the broadband slope of the spectrum. 

Given these limitations one may be tempted to simply use the linear Cold Dark Matter (CDM) power spectrum rather than the matter one (CDM + baryons). However, as shown by \citet{eisenstein98}, this has a different broadband slope which makes harder to highlight the BAO pattern. 

In order to better account for the broadband slope \citet{crocce08} define a smooth function by non-linearly evolving an initial wiggle-free spectrum using Renormalized Perturbation Theory. Here, we follow their approach and compute a smooth non-linear matter power spectrum from a $\Lambda$CDM-W7 simulation (see Table~\ref{cosmo}) with $2048^3$ particles
and $5250$~h$^{-1}$Mpc box (see Table~\ref{deusfurtest}) for which initial conditions have been generated using a wiggle-free spectrum from \citep{eisenstein98}. Upon doing so we have ensured that the effective one-dimensional amplitude of the large-scale velocity flow is the same as that inferred from the linear power spectrum computed with CAMB. The spectrum obtained from this simulation is then fitted with a polynomial of order 8 which defines our smooth power spectrum, $P_{\textrm{smooth}}^{\textrm{FUR}}(k)$ (see Appendix \ref{fit}). Here, the fact that we use a fit does not affect the determination of the BAO since we are directly fitting a smooth function. This is also the reason as to why $P_{\textrm{smooth}}^{\textrm{FUR}}(k)$ can be inferred without the need of running an extremely large volume simulation.

The wiggle-only spectrum of BAO is then given by
\begin{equation}\label{PBAO}
P_{\textrm{BAO}}(k)=P(k)-P_{\textrm{smooth}}^{\textrm{FUR}}(k),
\end{equation}
where $P$ is the matter power spectrum. Such definition solely depends on the choice of the initial smooth power spectrum that we have used to generate the initial conditions of the wiggle-free N-body simulation. Nevertheless, the conclusions depend only weakly on such choice since the initial wiggle-free spectrum given by \citet{eisenstein98} formula interpolates through the peaks and dips of the BAO at the initial redshift. In principle, one could opt to directly subtract the smooth spectrum computed from the N-body simulation rather than using its polynomial fit. This has the advantage of cancelling out statistical fluctuations at large sales. However, this may also introduce additional uncertainties at smaller scales due to the fact that non-linear effects cause mode couplings and thus the cancellation can lead to spurious effects. 

\begin{figure}
\begin{center}
\includegraphics[width=\hsize]{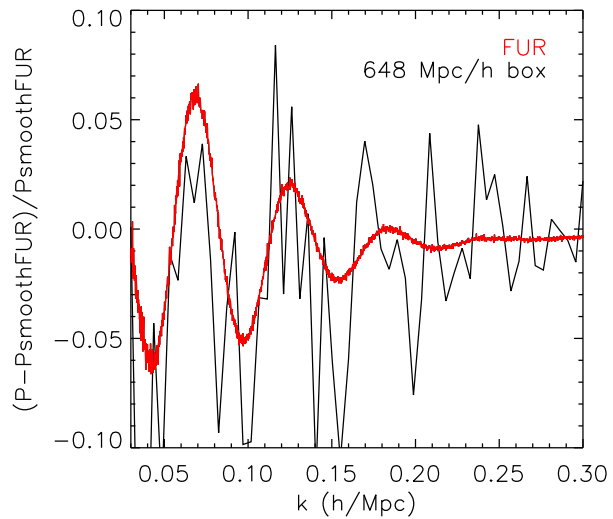}
\end{center}
\caption{Relative difference of the BAO spectrum with respect to the wiggle-free DEUS-FUR spectrum at $z=0$ for DEUS-FUR $\Lambda$CDM-W7 (red solid line) and in the case of a simulation with $L_\textrm{box}=648$~h$^{-1}$Mpc (black solid line). In the latter case large statistical fluctuations blur the BAO signal.} 
\label{boxsizebao}
\end{figure}

We plot in Figure~\ref{boxsizebao} the relative difference of the BAO spectrum with respect to the wiggle-free DEUS-FUR spectrum at $z=0$ for DEUS-FUR $\Lambda$CDM-W7 (red solid line) and in the case of a simulation with $648$~h$^{-1}$Mpc box length (black solid line). This illustrates the advantage of studying BAO with DEUS-FUR, since disposing of cosmic variance limited measurements allows us to finely resolve the BAO structure. Four peaks and troughs are clearly distinguishable, with the amplitude of the first oscillation (peak-to-trough) $\sim 13\%$, while that of the fourth one is of only half-percent. 

\subsection{BAO vs Semi-Analytical Model Predictions}
It is beyond the scope of this work to test against DEUS-FUR all existing semi-analytical models that aim to predict the non-linear corrections to the linear matter power spectrum. Instead, we focus on two such models that are exemplary of substantially different approaches, one based on a perturbative calculation and the other on a combination of fit to simulations and the halo model. The former can provide accurate predictions on quasi-linear scales, while failing deep in the non-linear regime \citep[see e.g.][]{carlson09}. The latter, while more accurate at small scales, is less at intermediate ones (0.1-0.3~h Mpc$^{-1}$). In the first case we consider the two loop regularized multi-point propagator method (RegPT) for which we compute the non-linear matter power spectrum using the code RegPT \citep{taruya12}, while in the other case we consider the widely used fitting prescription Halofit \citep{smith03}.

\begin{figure*}
\begin{tabular}{cc}
\includegraphics[width=0.45\hsize]{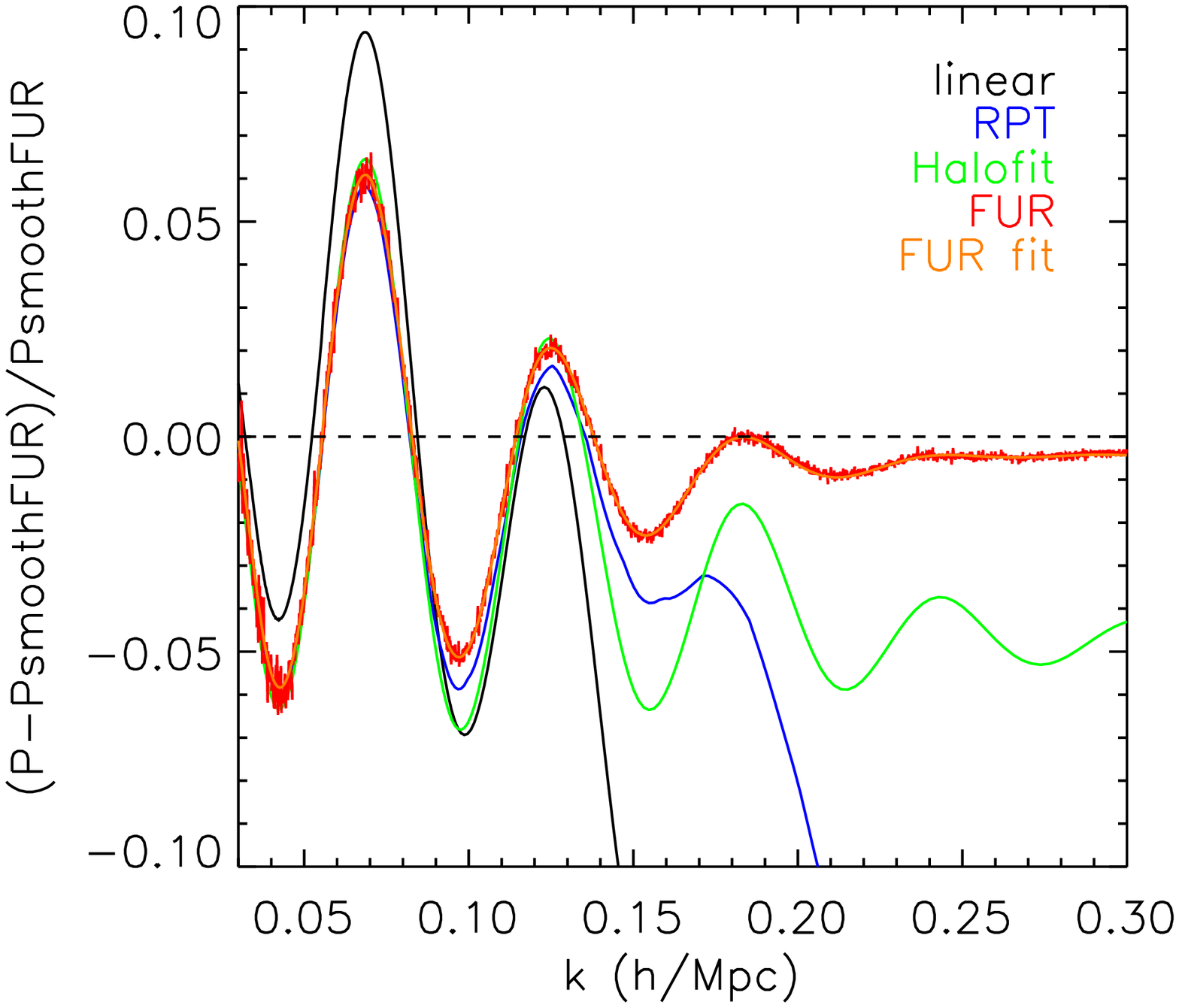}&\includegraphics[width=0.45\hsize]{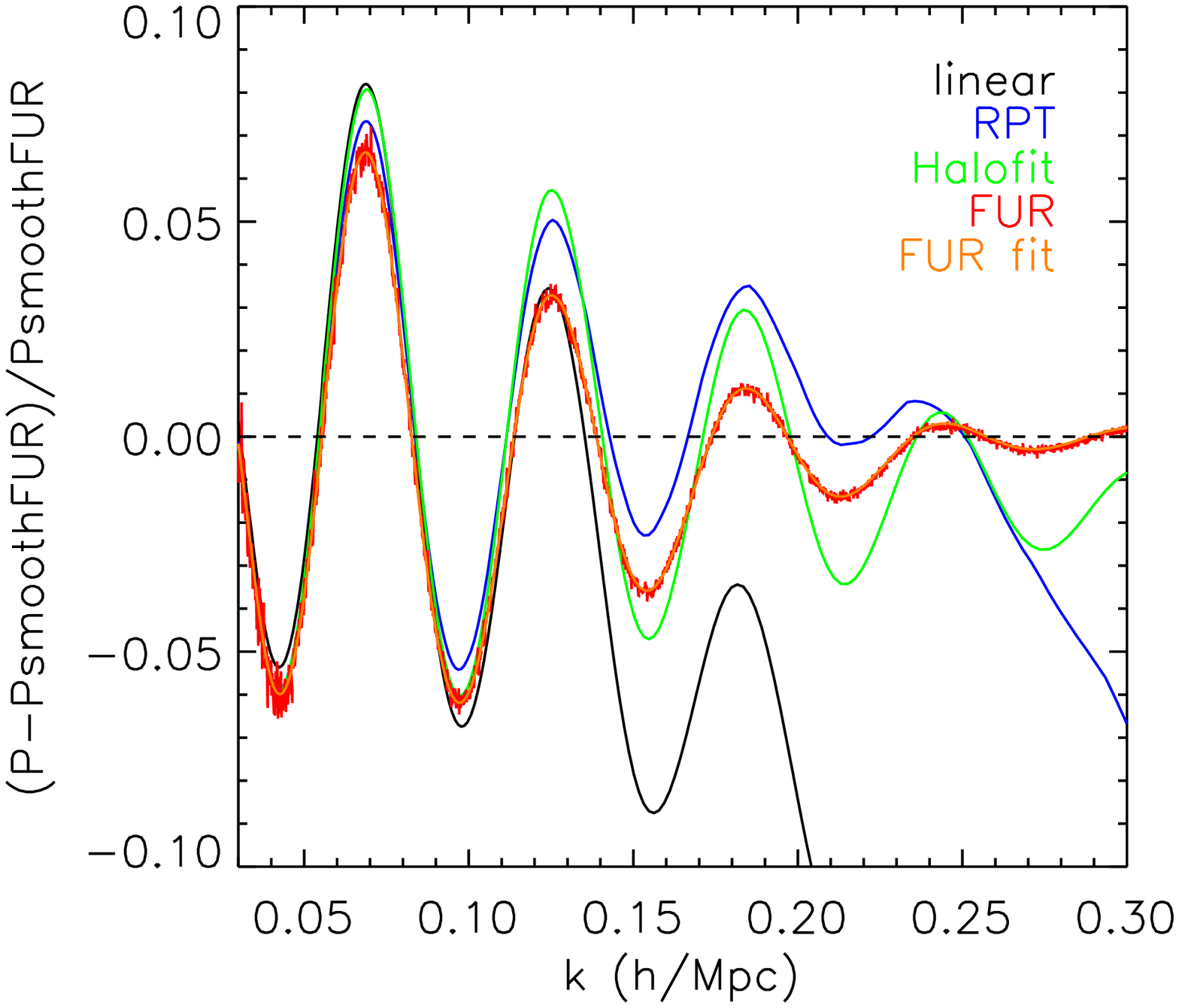}\\
\includegraphics[width=0.45\hsize]{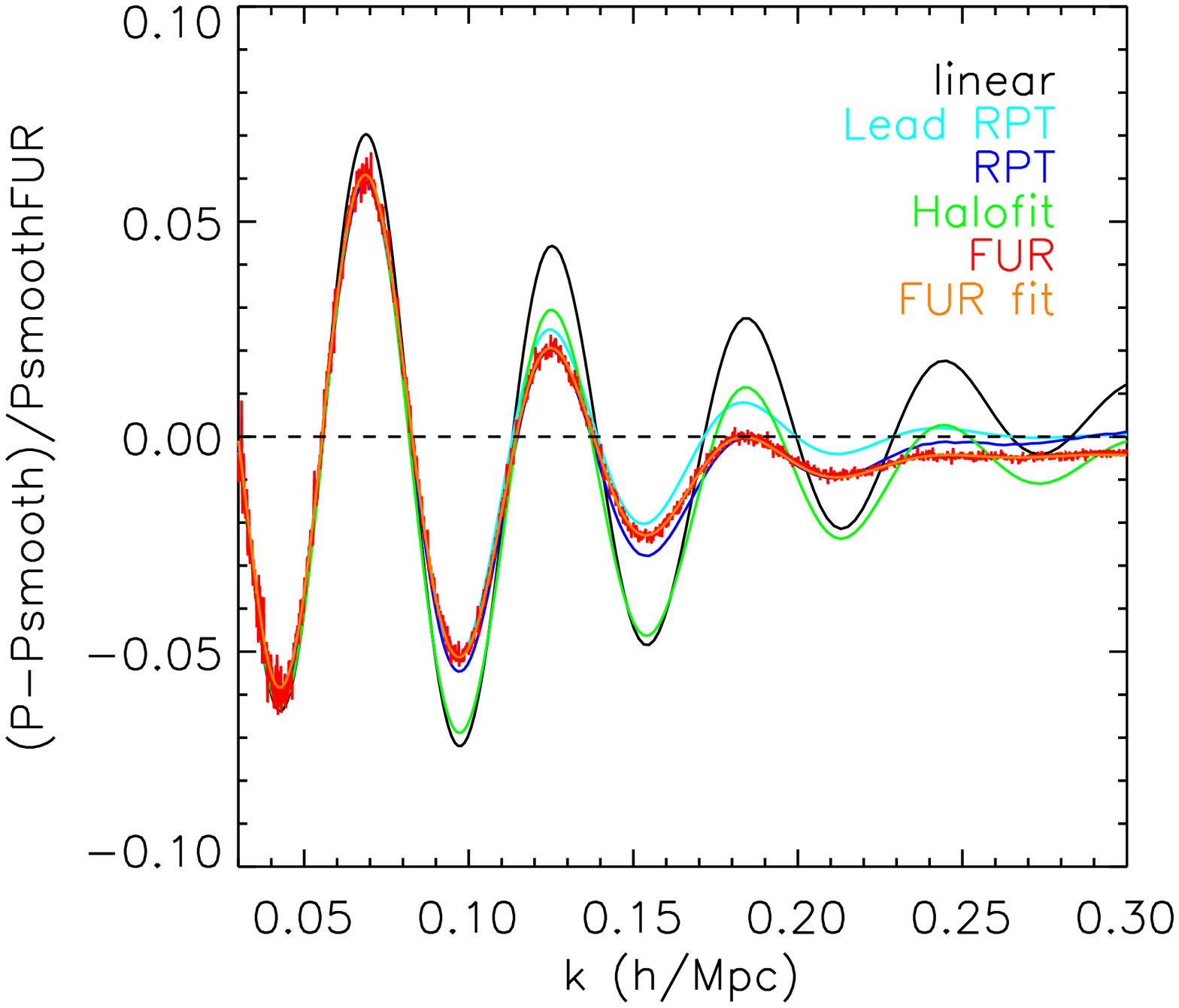}&\includegraphics[width=0.45\hsize]{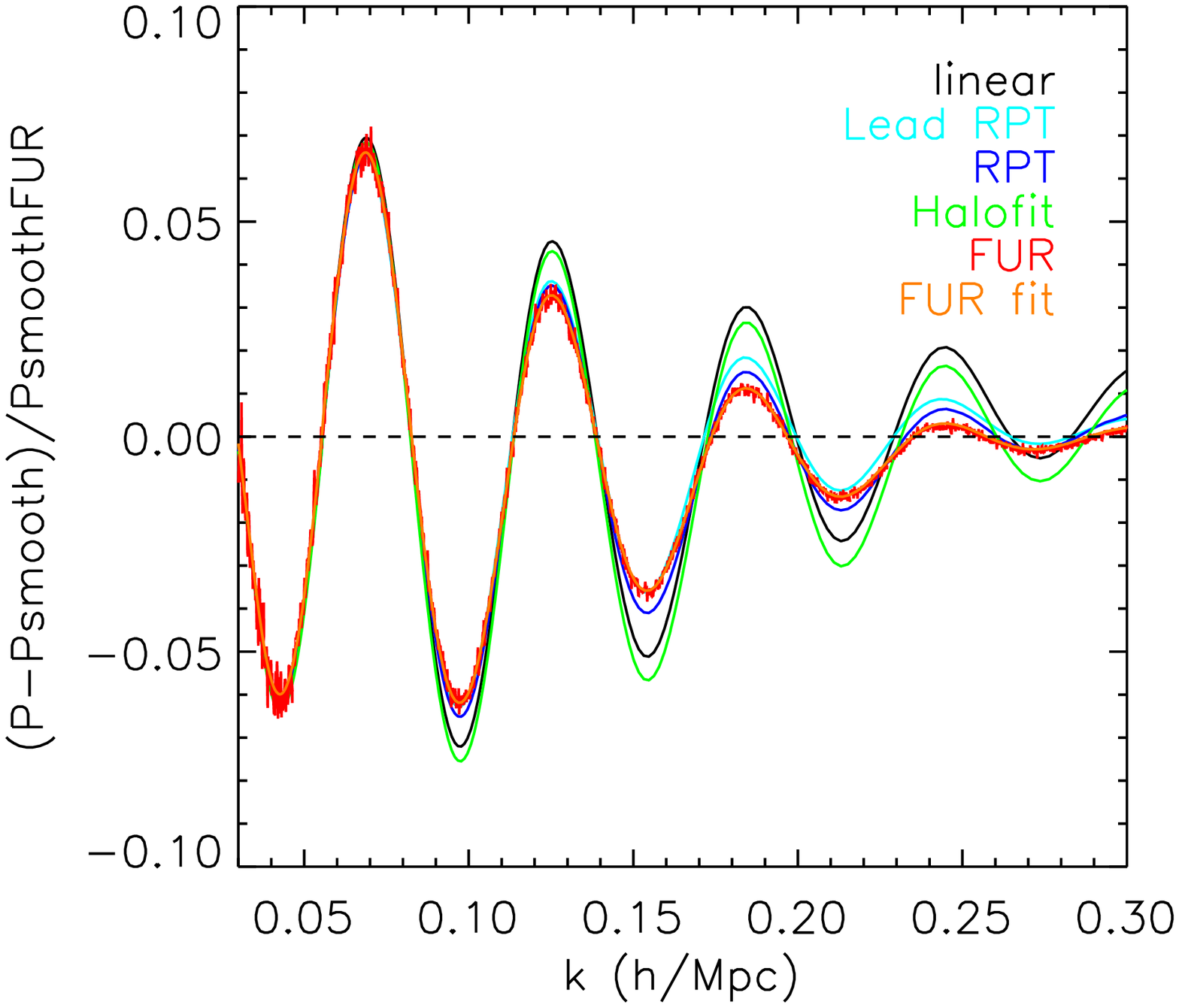}\\
\end{tabular}
\caption{Top-panels: Relative difference of the BAO power spectrum to $P_{\textrm{smooth}}^{\textrm{FUR}}(k)$ at $z=0$ (left panel) and $1$ (right panel). The different lines correspond to the linear prediction (black line), RegPT (blue line), Halofit (green line), DEUS-FUR (red line) and DEUS-FUR fit (orange line) respectively. Bottom-panels: Difference with respect to the smooth spectrum of each model normalized to $P_{\textrm{smooth}}^{\textrm{FUR}}(k)$ at $z=0$ (left panel) and $1$ (right panel). In addition to the models shown in the top-panels we include the case of the leading term from Renormalized Perturbation Theory (light blue line).}
\label{analytical}
\end{figure*}

The results of the comparison are summarized in Figure~\ref{analytical}. In the top panels we plot the relative difference of the matter power spectrum with respect to the DEUS-FUR smooth spectrum at $z=0$ (left panel) and $z=1$ (right panel) respectively. The different lines correspond to the linear prediction (black solid line), RegPT (blue solid line), Halofit (green solid line), DEUS-FUR (red solid line) and DEUS-FUR fit (orange solid line, see Appendix \ref{fit}). First, it is worth noticing that non-linear modifications of the matter power spectrum occur over the entire BAO range. At $z=0$ (top-left panel) the discrepancy between the linear theory and DEUS-FUR is of $\sim 2\%$ on the first trough and $\sim 4\%$ on the first peak, while at $z=1$ (top right panel) this reduces to $\sim 1\%$ and $\sim 2\%$ respectively, but still above cosmic variance errors. We may also notice that RegPT and Halofit are in agreement with DEUS-FUR to within $1\%$ on the first trough at $z=0$, while larger deviations occur at smaller scales. In particular, Halofit deviates at $1\%$ level on the first peak and underestimates the damping of the BAO up to $4\%$ beyond the second peak, while it recovers to good approximation the position of the extrema. RegPT shows deviation of order $1\%$ up to the second peak, but afterward it rapidly diverges. At $z=1$ the non-linear damping of the BAO is smaller, consequently the fourth peak has become more prominent, while even the fifth trough has appeared. Deviations from the linear prediction are now shifted to larger $k$, with the linear amplitude being off by $\gtrsim 10\%$ beyond the third trough. The maximal deviation of Halofit and RegPT is $\sim 2\%-3\%$, with RegPT diverging beyond the fourth peak.

The discrepancy of the semi-analytic models with respect to DEUS-FUR is largely due to the broadband slope of the non-linear power spectrum. In the bottom panels of Figure~\ref{analytical} we plot the BAO spectrum obtained by subtracting the smooth spectrum predicted by each model. In this case we also plot the prediction from the leading term of Renormalized Perturbation Theory (lRPT) given by
\begin{equation}\label{lRPT}
P^{\textrm{lRPT}}_{\textrm{BAO}}(k)=\left[P_{\textrm{lin}}(k)-P^{\textrm{lin}}_{\textrm{smooth}}(k)\right] \times \exp{\left(-\frac{ \sigma^2_v k^2}{2}\right)},
\end{equation}
with $P_{\textrm{lin}}(k)$ the linear power spectrum, $P^{\textrm{lin}}_{\textrm{smooth}}(k)$ the smooth (wiggle-free) linear power spectrum and 
\begin{equation}
\sigma_v^2=\frac{1}{3 \pi^2} \int P_{\textrm{lin}}(k)dk 
\end{equation}
is the effective one-dimensional amplitude of large-scale velocity flows \citep{crocce08}. This expression coincides with that from the peak-background split to lower order as well as the resummed Lagrangian Perturbation Theory, as such it has motivated several fitting formulae of the BAO. 

We notice that Halofit now has deviations of order of $2-3\%$ on dips and $1-2\%$ on peaks at $k>0.08$~h Mpc$^{-1}$ both at $z=0$ and $1$ respectively. 

RegPT shows a remarkable agreement with DEUS-FUR with differences $\lesssim 1\%$ up to the third trough, which is well beyond the supposed range of validity of RegPT. Slightly larger deviations occurs only beyond the third trough at $z=0$. The case of the leading RPT term (light blue line) also provides a good description to the DEUS-FUR spectrum only a few percent worse than RegPT.

Next, we will perform a detailed quantitative analysis of the non-linear effects on the BAO extrema.

\begin{figure}[th]
\begin{center}
\includegraphics[width=\hsize]{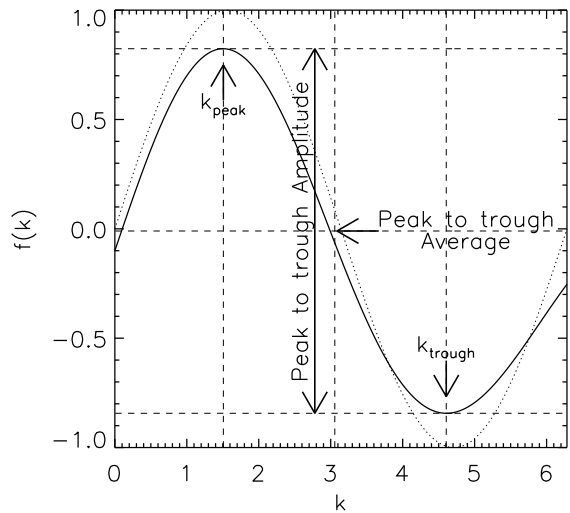}
\end{center}
\caption{We decompose an oscillatory pattern in $k$-space into the position of the extrema, $k_{\textrm{peak}}$ and $k_{\textrm{trough}}$ for peak and trough respectively, the peak-to-trough amplitude (long vertical arrow) and the peak-to-trough average (horizontal arrow). The dotted curve represents an unperturbed oscillatory function (sine) before application of an artificial shift, damping and modification of the broadband slope (solid curve). The difference between the horizontal positions of the extrema in the modified and original signal is called ``shift'', the ratio of the peak-to-trough between the modified and original curve is called ``damping'', while the difference of the two averages gives information on the broadband slope of the perturbed curve. We use this decomposition to characterize the BAO pattern by directly measuring these quantities in the BAO DEUS-FUR spectrum. These should not be confused with the quantities entering in the fiting formula Eq.~(\ref{eqfit}).}  
\label{peak_to_trough}
\end{figure}

%%%%%%%%%%%%%%%%%%%%% SECTION 4 %%%%%%%%%%%%%%%%%%%%%%%%%%%%%%%%%%%%%%%%%%%%%%%%

\section{Featuring BAO pattern}\label{BAO_NL}

The non-linear clustering of matter shifts the position of the BAO peaks and dips, damps their amplitude and alters the broadband slope of the power spectrum. These are not independent effects, since the damping and the broadband slope both contribute to shifting the position of the BAO extrema relative to the linear case. Nevertheless such decomposition provides a phenomenologically meaningful way of quantifying the effect of non-linearities, testing the accuracy of semi-analytical model predictions or comparing them against observational measurements (to this purpose we provide in Appendix~\ref{fit} a fitting formula of the DEUS-FUR spectrum). 

Non-linearities are usually considered to be a nuisance which degrades the cosmic distance information encoded in the BAO. We will show that an accurate modelling of these non-linear features can provide additional cosmological information as they probe the linear growth rate of cosmic structures.

The non-linear shift of the position of the BAO extrema has been quantified using perturbation theory in \citet{nishimichi07}. Here, we perform a detailed numerical analysis using the BAO spectrum from DEUS-FUR. Our aim is to evaluate in absolute terms the effect of non-linearities on the characteristics of BAO extrema, hence differently from the previous Section we normalize the BAO spectrum to the amplitude of the DEUS-FUR spectrum at $z=0$ on a linear scale, $P_{\textrm BAO}^{\textrm FUR}(k=0.1)$, such as not to alter the extrema.

As already stressed, differently from previous studies \citep[see e.g.][]{seo05,angulo08,seo08,seo10}, we will characterize each BAO extremum directly from the measured N-body spectrum, rather than a best-fitting function. We detect the BAO peaks and dips performing a local second-order least-square polynomial fit (Savitsky-Golay filter) of width 100 data point which takes into account DEUS-FUR error bars. In Figure~\ref{peak_to_trough} we illustrate our decomposition of the BAO extrema in terms of the horizontal position, peak-to-trough amplitude and peak-to-trough average.

In order to consistently compare with the semi-analytical model predictions we have binned the corresponding power spectra as the DEUS-FUR case and applied the same procedure to estimate the location and amplitude of the BAO extrema. 

\begin{figure*}
\begin{center}
\includegraphics[width=\hsize]{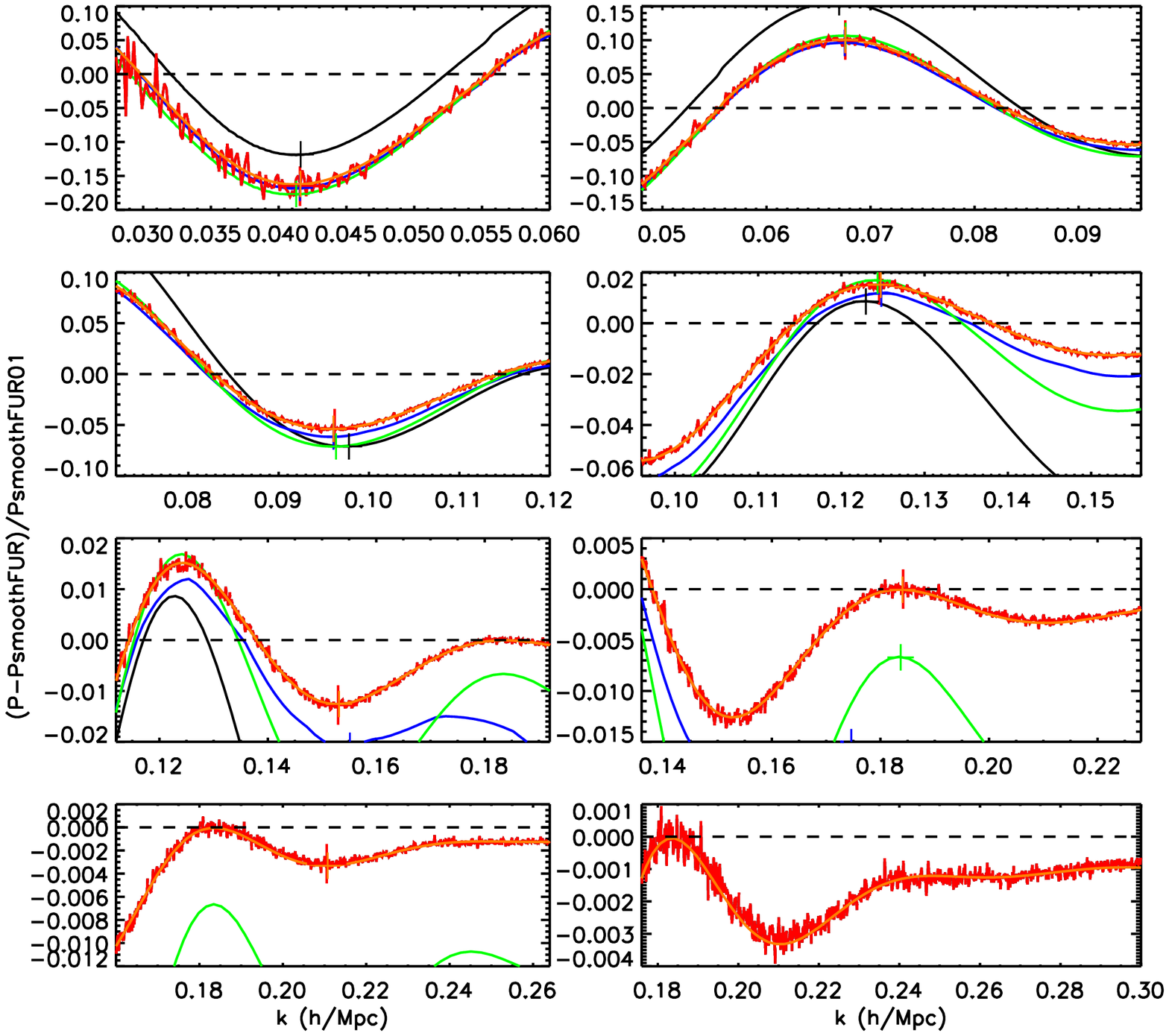}
\end{center}
\caption{Zoom on the peaks and dips of the BAO spectrum at $z=0$ normalized to the smooth DEUS-FUR power spectrum at $k=0.1$~h Mpc$^{-1}$. The different lines correspond to semi-analytic model predictions plotted in Figure~\ref{analytical}. Crosses mark the position of the extrema.}  
\label{extremaFUR}
\end{figure*}

\begin{figure*}
\begin{center}
\includegraphics[width=\hsize]{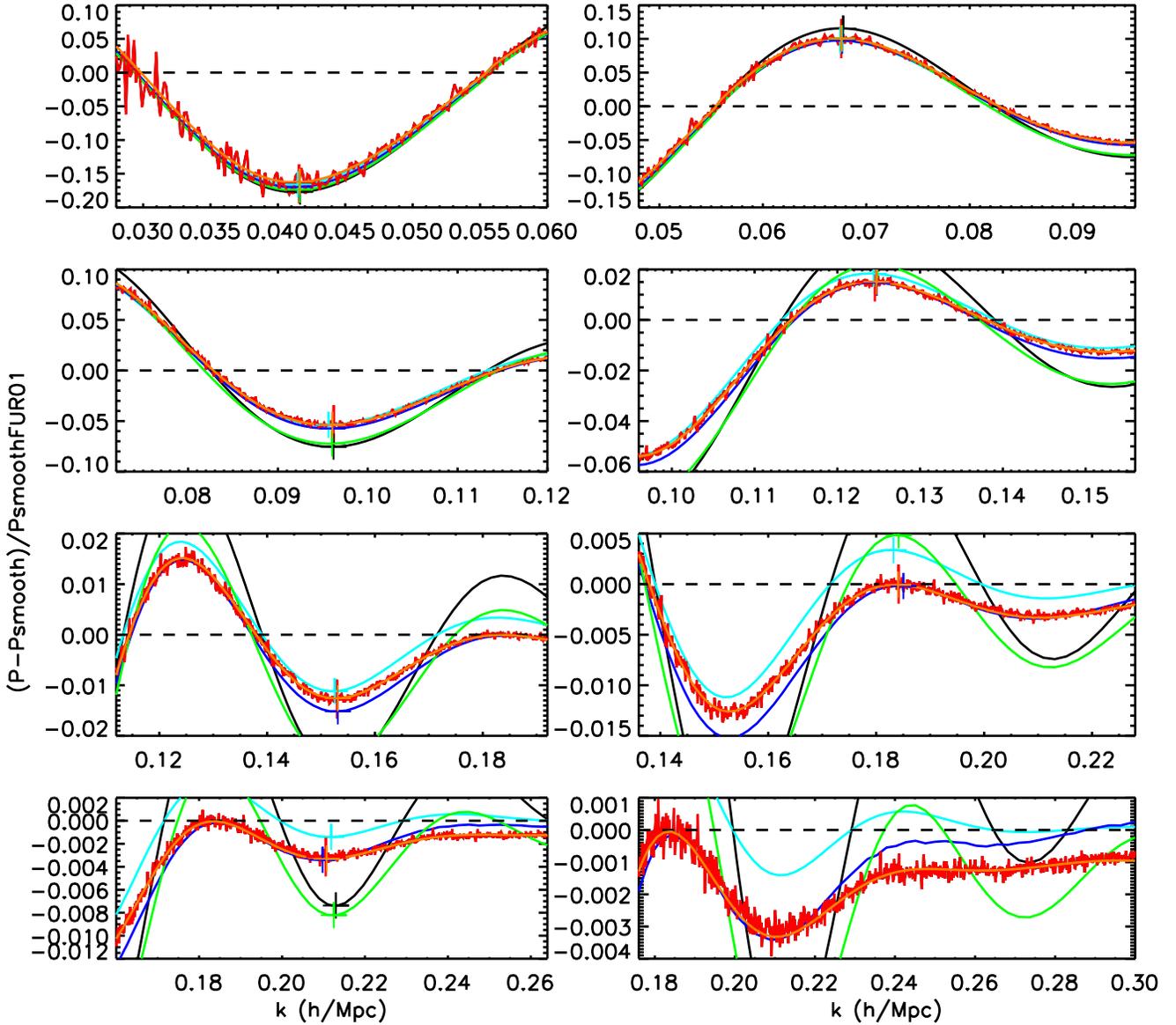}
\end{center}
\caption{As in Figure \ref{extremaFUR} after subtraction of the smooth power spectrum predicted by each model.}  
\label{extrema}
\end{figure*}

In Figure~\ref{extremaFUR} we plot a zoom on the first four troughs (left panels) and peaks (right panels) from DEUS-FUR (red line), linear theory (black line), RegPT (blue line) and Halofit (green line) respectively. Crosses mark the location of the detected extrema. In Figure~\ref{extrema} we plot the same zoom on the extrema having subtracted the smooth power spectrum predicted by each model.

\begin{figure*}
\begin{tabular}{cc}
\includegraphics[width=0.4\hsize]{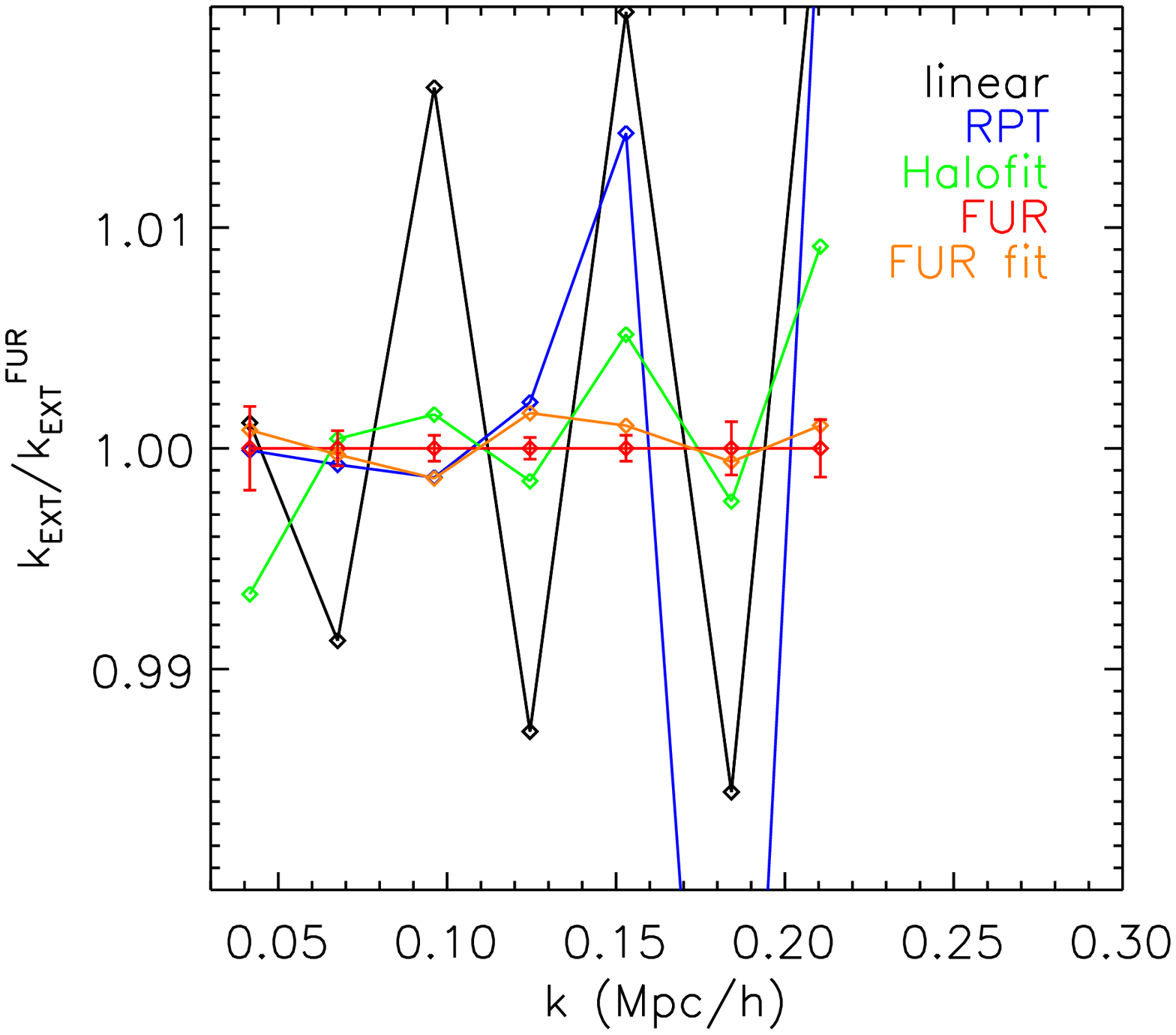}&\includegraphics[width=0.4\hsize]{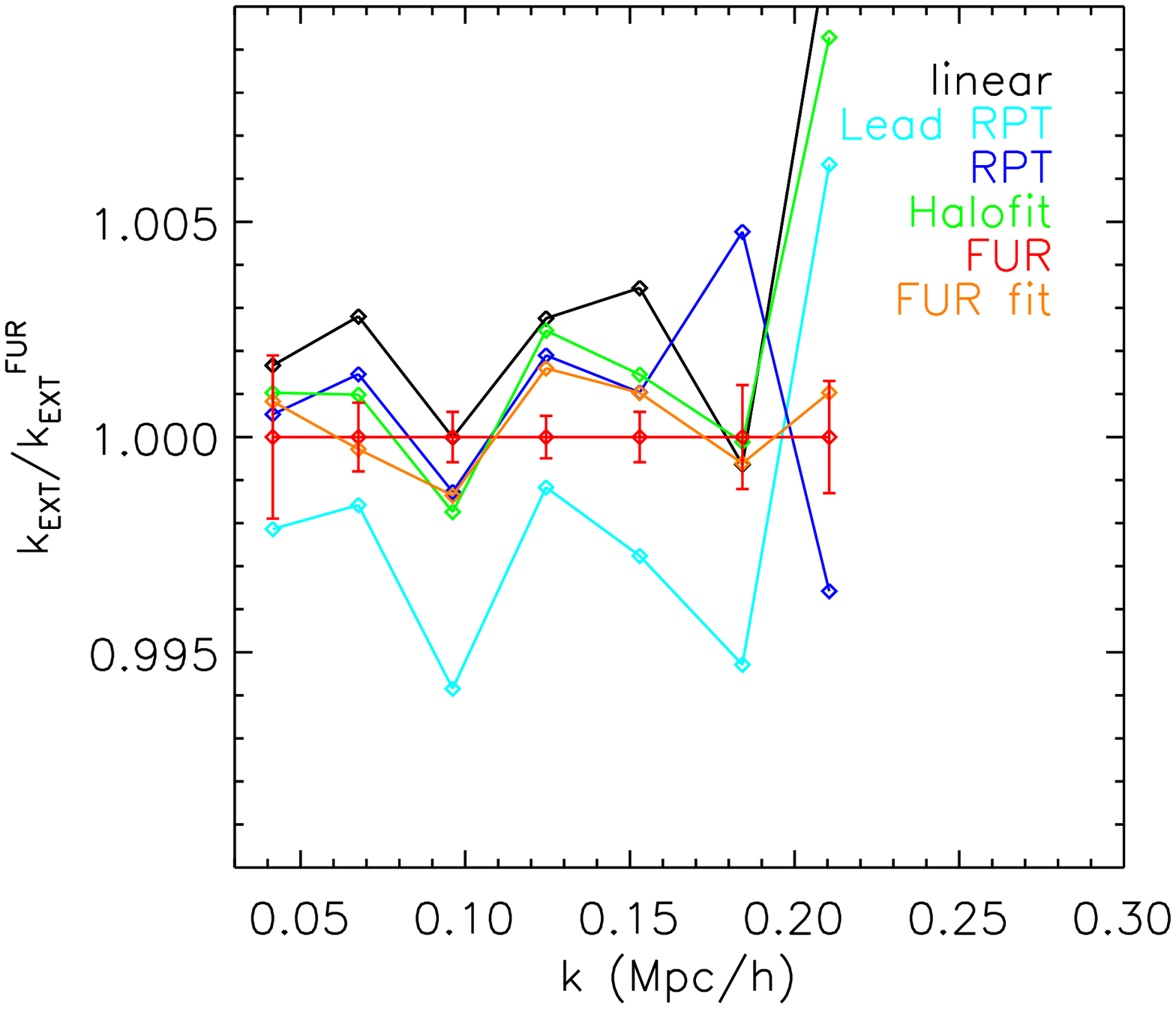}\\
\includegraphics[width=0.4\hsize]{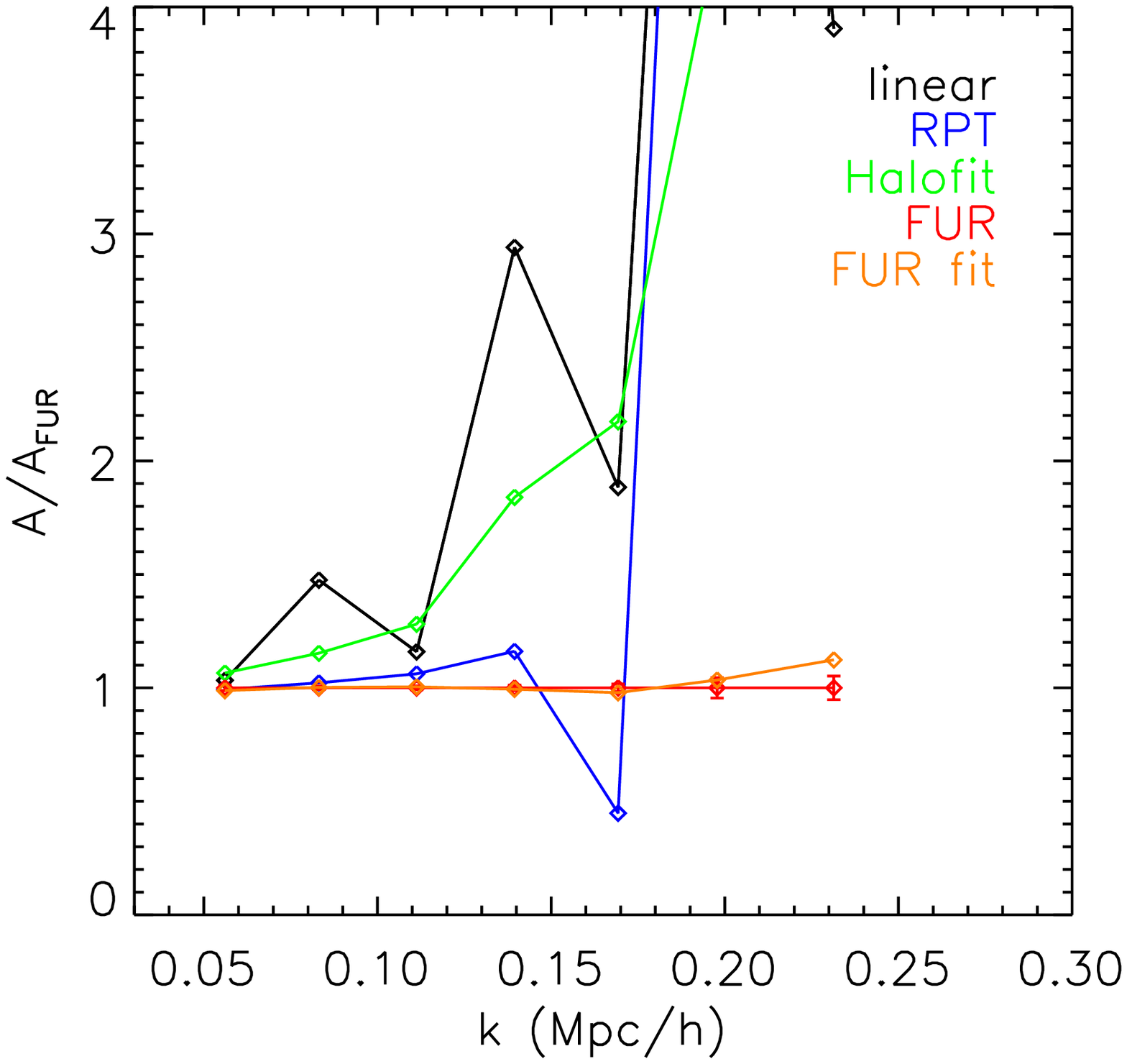}&\includegraphics[width=0.4\hsize]{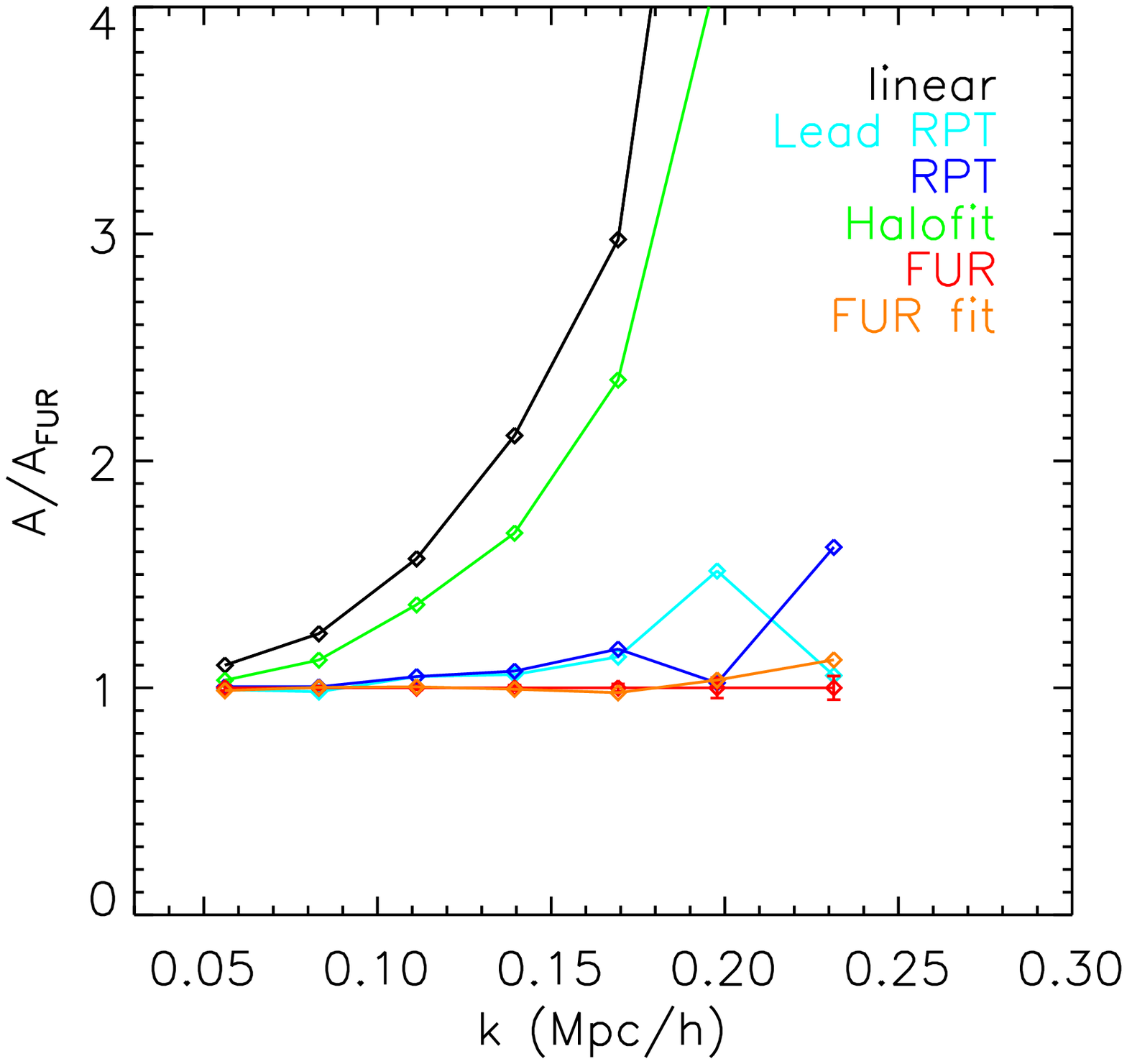}\\
\includegraphics[width=0.4\hsize]{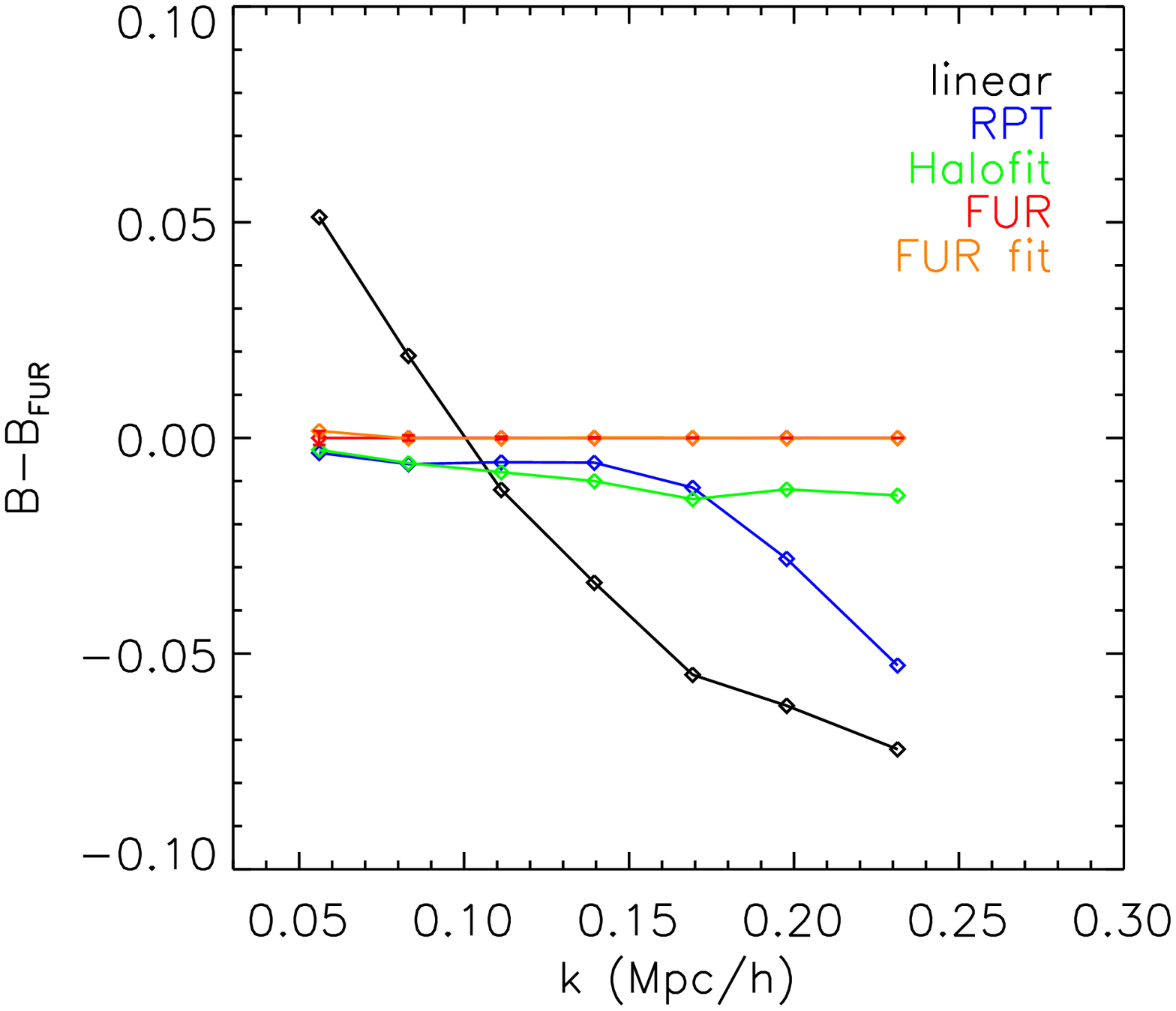}&\includegraphics[width=0.4\hsize]{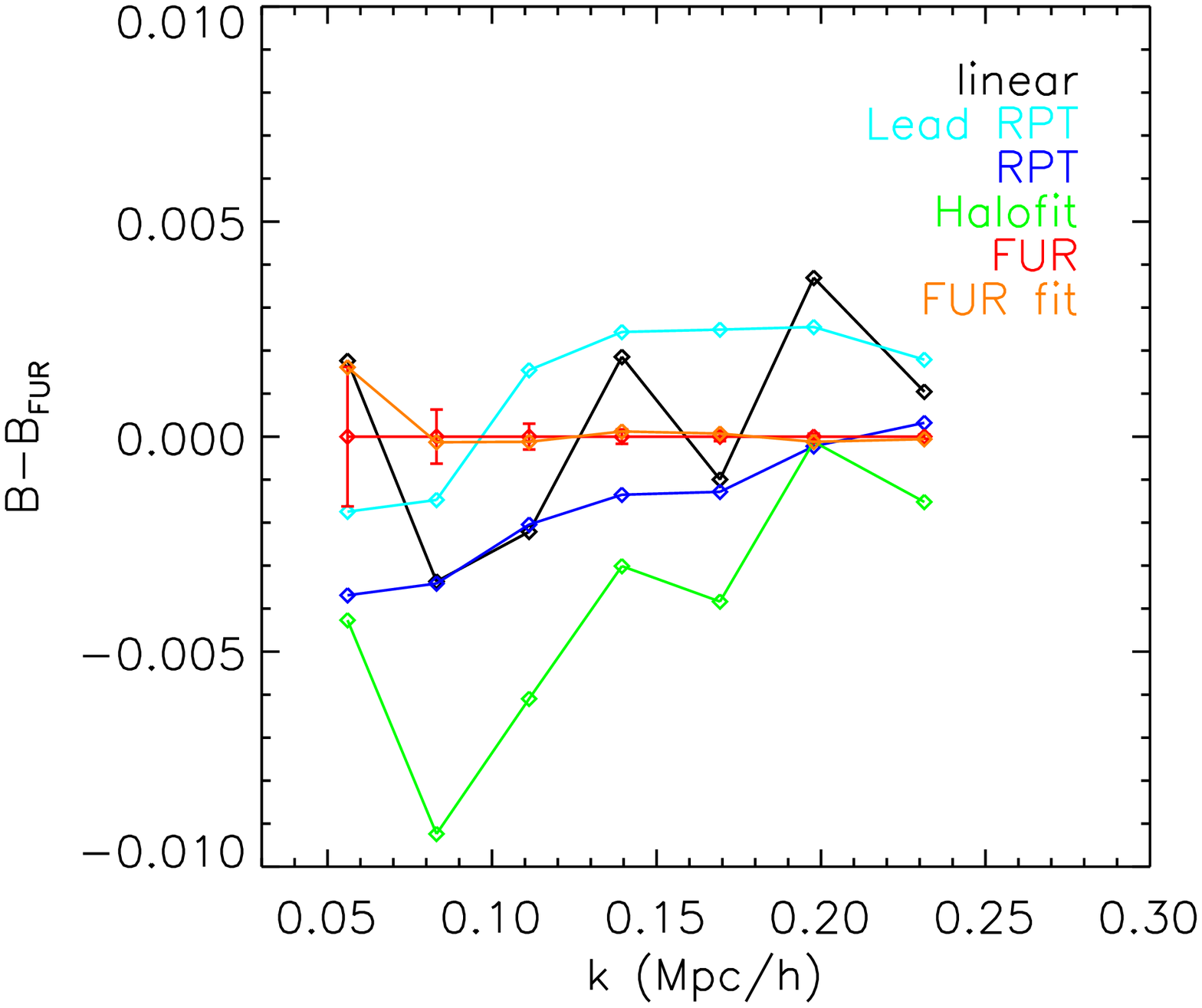}\\
\end{tabular}
\caption{Characteristics of the BAO extrema for different models with respect to the DEUS-FUR BAO spectrum at $z=0$. The left panels show the case of a BAO spectrum defined with respect to the wiggle-free DEUS-FUR spectrum, while the right panels show the case in which the smooth spectrum predicted by each model has been subtracted. Top panels: position of the BAO model extrema relative to that of DEUS-FUR BAO. This is related to the shift of BAO extrema. Middle panels: relative amplitude of the neighbouring pairs of extrema in BAO model ($A$) relative to that of DEUS-FUR BAO ($A_{\textrm{FUR}}$). This is related to the damping of BAO amplitude. Bottom panels: difference of the average between neighbouring extrema in BAO model and DEUS-FUR BAO. This is related to the coupling between BAO signal and the broadband slope. The different lines correspond to models as in Figure~\ref{analytical}.}  
\label{physics}
\end{figure*}

\subsection{Shift of BAO extrema}
The physical origin of this shift is discussed in several papers \citep{eisenstein07,nishimichi07,crocce08,padmanabhan09,sherwin12}. An accurate determination of the non-linear shift of BAO extrema is crucial to infer unbiased cosmic distance measurements. As shown in \citep{angulo08} an error of $1\%$ on the peak position at $z=1$ leads to a $4\%$ error on a Dark Energy equation of state parameter (assuming the other cosmological parameters to be known). 

In the top-left panel of Figure~\ref{physics} we plot the shift of the BAO peaks (even points) and dips (odd points) at $z=0$ defined as the ratio between the location of the extrema from the linear theory (black line), RegPT (blue line), Halofit (green line) relative to DEUS-FUR (red line). We can see that the linear theory misestimates the location of the extrema at $>1\%$ beyond the first dip. In contrast, RegPT deviates at more than $1\%$ only beyond the third dip as consequence of the drop of the broadband slope. Halofit performs better with deviations $<1\%$ up to the fourth dip. 

Subtracting the wiggle-free spectrum predicted by each of the models greatly reduces the discrepancies. This can be seen in the top-right panel of Figure~\ref{physics}. In the case of RegPT the deviations are well within $1\%$ up to the fourth dip, similarly for Halofit and the leading RPT term. Even the linear theory is within $1\%$ up to the third peak. 

We now focus on the redshift dependence of the shift of the position of each BAO extrema in DEUS-FUR with respect to that in the linearly evolved initial power spectrum. This is plotted for different extrema in the left panel of Figure~\ref{evol}. We can see that the amplitude of the shift is $\lesssim 1\%$ at all redshifts. It is worth noticing that the second dip and the third peak remains unaltered by the non-linear effects with shifts which are $<0.1\%$. Given the fact that the latest BAO measurements already test the shift of the BAO at a few percent level \citep{anderson2012,padmanabhan12,xu2012}, this result suggests that upcoming cosmic distance measurements from BAO may derive unbiased estimates by specifically focusing on these two extrema. We also remark that the redshift dependence is well described by the square of the linear growth function $D_+^2(z)$ (dashed line) as expected from perturbation theory \citep{crocce08,padmanabhan09,seo10}. Hence, future surveys may have enough resolution to detect such a trend and independently probe the linear growth rate. 

\subsection{Damping of BAO amplitude}
The suppression of BAO is the most visible non-linear effect which is direct consequence of the relative displacement (5-10~h$^{-1}$Mpc) of pair of particles separated by $\sim 100$~h$^{-1}$Mpc \citep{eisenstein07}. Because of this, a measurement of the BAO damping factor can in principle probe the lagrangian displacement field and provide constraints on the growth rate of cosmic structures \citep[see e.g.][]{nomura08,nomura09}.

We estimate the damping of the BAO by measuring the absolute difference of the amplitude of two consecutive extrema (i.e. first trough to first peak, first peak to second trough,...). We plot the predicted amplitudes for the different models relative to DEUS-FUR case in the middle-left panel of Figure~\ref{physics}. We can see that both the linear theory and Halofit overestimate the amplitude of the extrema with deviations exceeding $50\%$ at $k>0.11$~h Mpc$^{-1}$. In the right panel we show the same amplitudes estimated after subtracting the wiggle-free spectrum predicted by each model. In such a case we can clearly see that the linear calculation and Halofit exponentially overestimate the amplitude of the BAO extrema. In the case of RegPT the discrepancy amounts to $15\%$ up to $k=0.14$~h Mpc$^{-1}$ and diverges afterward, whereas it reduces to no more than $10\%$ up to $k=0.20$~h Mpc$^{-1}$ after subtracting the smooth component and increases up to $50\%$ on the extrema at the highest $k$. 

We now estimate the damping factor in DEUS-FUR BAO spectrum relative to the linear prediction. This is defined as
\begin{equation}
\Sigma=\sqrt{-\frac{4}{k_p^2+k_t^2}\ln{\left(\frac{A_{\textrm FUR}}{A_{\textrm LIN}}\right)}},
\end{equation}
where $A_{\textrm FUR}$ and $A_{\textrm LIN}$ are the peak to trough amplitude in DEUS-FUR and the linear theory (initial condition) respectively, $k_p$ and $k_t$ are the wavenumbers of the peak and trough respectively. The above formula converges to first order to the case of an exponential term $\propto \exp (-\Sigma^2 k^2/2 )$ with constant damping rate $\Sigma$. 

\begin{figure*}
\begin{tabular}{cc}
\includegraphics[width=0.45\hsize]{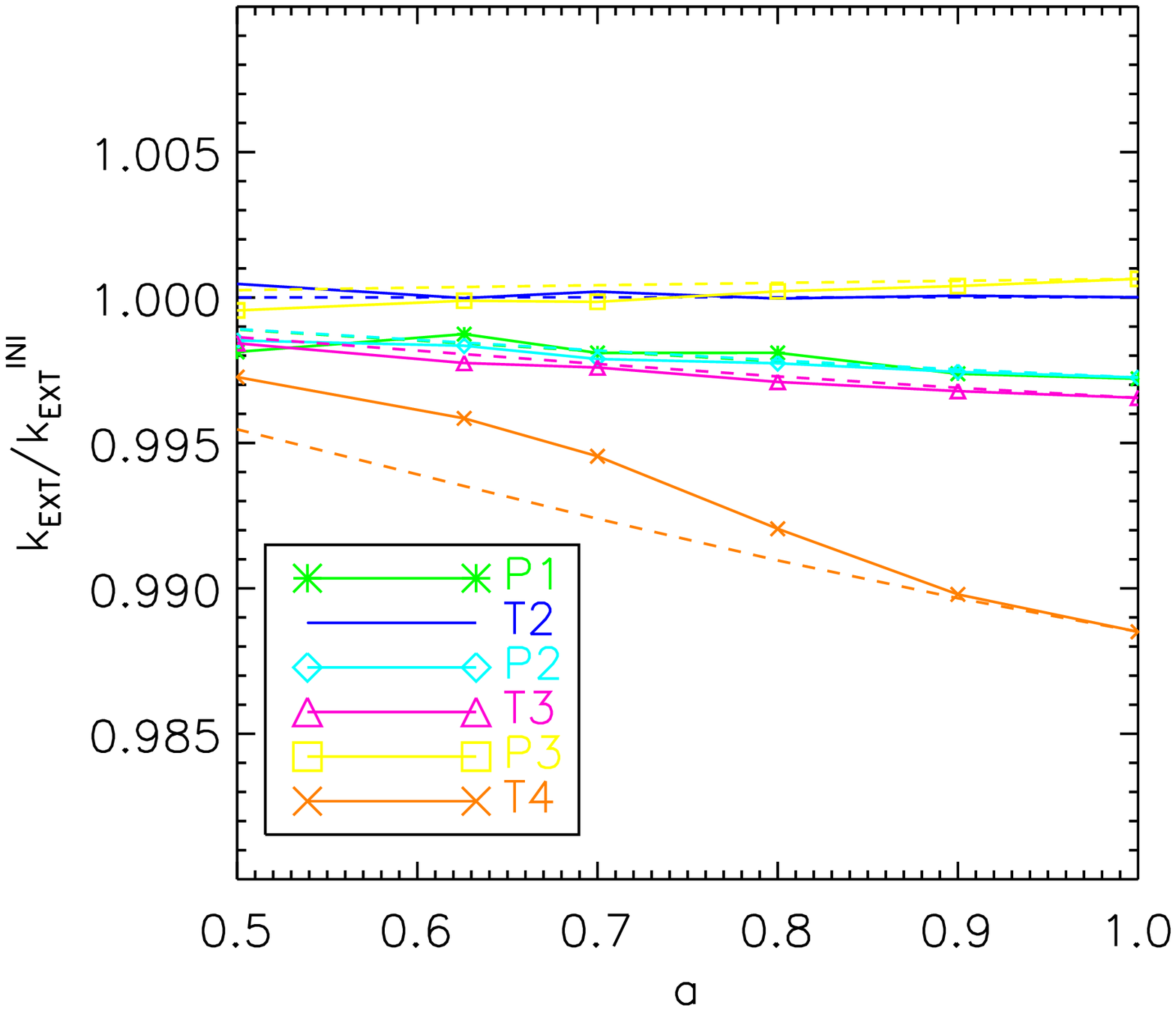}&\includegraphics[width=0.45\hsize]{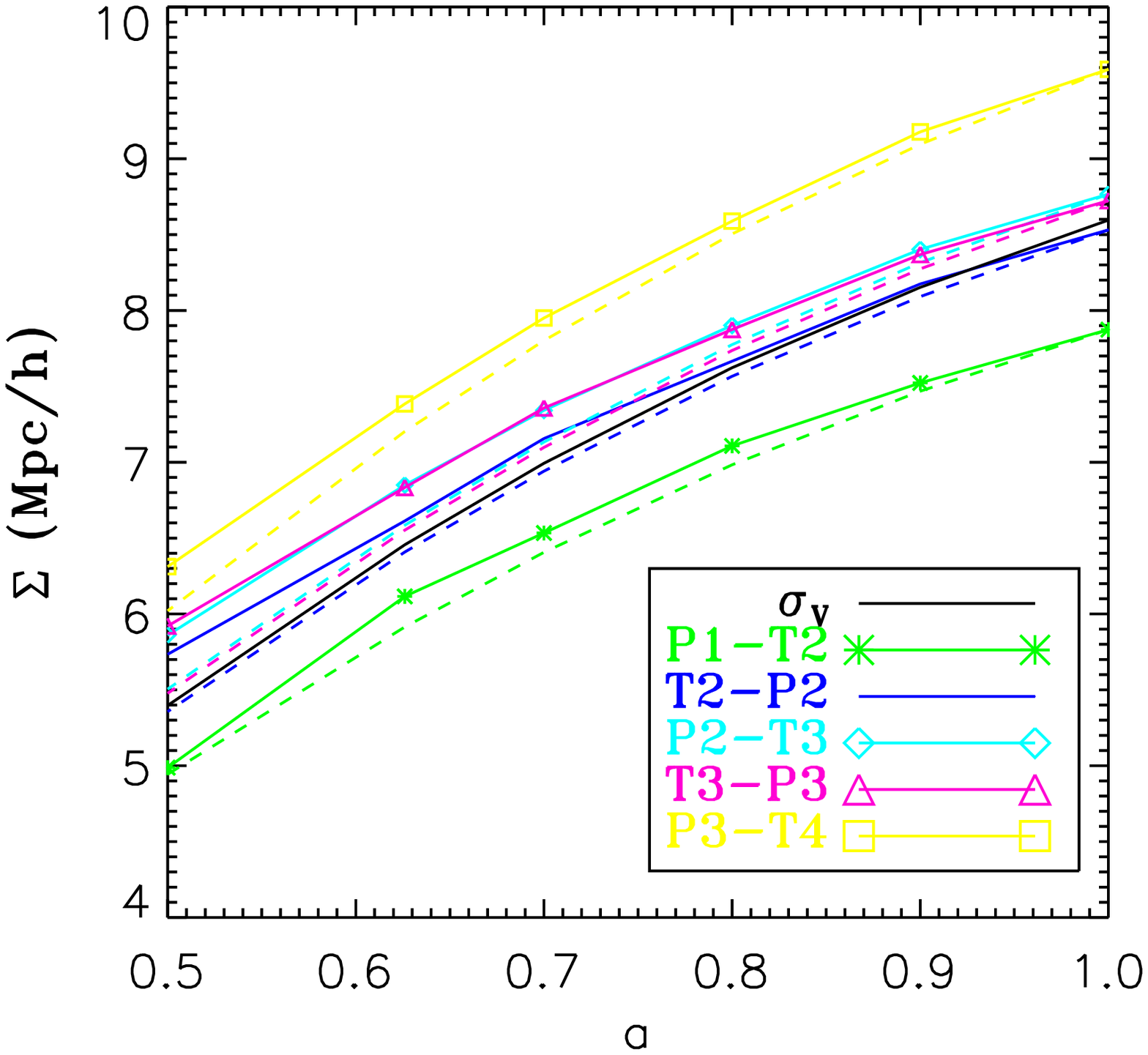}\\
\end{tabular}
\caption{Left panel: evolution of the shift of BAO extrema as function of the scale factor $a$. The different solid lines correspond to peaks (P) and troughs (T) as in the legend, while the dashed lines show the square of the linear growth function $D_+^2(a)$ from perturbation theory. Right panel: evolution of the damping factor of BAO extrema as function of the scale factor. The black solid line shows the evolution of $\sigma_v(a)$, while the dashed lines corresponds to $D_+(a)$ scaling of the damping of the BAO extrema as expected from perturbation theory.} 
\label{evol}
\end{figure*}

In the right panel of Figure~\ref{evol} we plot $\Sigma$ as function of redshift for the BAO extrema in DEUS-FUR. First, we can see that the damping does not behave as a constant term $\Sigma$, rather it increases as function of $k$ from one extrema to the next. Secondly, it decreases as function of redshift as expected from perturbation theory. However, the exact value for each pair of extrema is not given by $\sigma_v$ as expected from Eq.~(\ref{lRPT}). 
Because of these features a global fit to the BAO would lead to a different estimation of the damping factor depending on the choice of the fitting range or the relative size of experimental errors. On the other hand, we can see that the redshift evolution of the damping rate approximately scales as $\Sigma(z) \propto D^+(z)$ (dashed line). Thus, it should be possible to infer the linear growth rate from the measurement of BAO amplitude at two different redshifts.

\subsection{BAO coupling to broadband slope}
The inaccuracy in reproducing the broadband slope of the non-linear matter power spectrum is primarily responsible for the discrepancies between the semi-analytic model predictions and the DEUS-FUR spectrum. That is why subtracting the smooth wiggle-free spectrum predicted by each of the models provides a much better description. We quantify this effect by computing the average amplitude of neighbouring pairs of extrema. We plot the prediction for different models at $z=0$ relative to DEUS-FUR in the bottom-left panel of Figure~\ref{physics}. We can see that the linear theory overestimates the average amplitude up to $5\%$ at $k\lesssim 0.1$~h Mpc$^{-1}$, while it underestimates at larger $k$. Halofit and RegPT are in agreement with DEUS-FUR at $1\%$ level up to $0.15$~h Mpc$^{-1}$. At larger wavenumbers Halofit remains within $2\%$, while the prediction from RegPT sharply falls to more than $5\%$ above $0.2$~h Mpc$^{-1}$. Subtracting the wiggle-free spectrum of each model absorbs most of the broadband slope effect as shown in the bottom-right panel of Figure~\ref{physics}. Discrepancies are now $<0.5\%$, however it is worth noticing a mode dependent trends which differ from one model to another. This indicates the presence of a residual coupling between the BAO and the continuum which complicate the modelling of the non-linear effects on the BAO. That is why defining a wiggle-free spectrum through a fit to the observed spectrum may introduce spurious effects at the percent level.

\section{Conclusion}\label{conclu}
We have presented a study of the BAO in the real-space matter power spectrum from the DEUS-FUR simulation of $\Lambda$CDM-W7 cosmology. The simulation box covers the size of the full observable volume of the simulated cosmology thus enabling a cosmic variance limited estimation of the matter power spectrum. Using a testbed of large volume high-resolution simulations we are able to control numerical systematic errors to within $1\%$. This allows us to investigate for the first time the full BAO range ($0.03<k~[\textrm{h~Mpc}^{-1}]<0.3$) at low redshift (from $z=0$ to $z=1$ ) with less than $1\%$ systematic and statistical errors. 

Previous works dedicated to assessing the effect of non-linearities on the BAO have either relied on perturbation theory (which eventually breaks down at large wavenumbers) or used N-body simulations of smaller effective volumes without assessing the effect of numerical systematic errors. Furthermore, differently from previous studies, the very narrow sampling in $k$ (with bins of size $\Delta k/k\lesssim 1\%$) accessible through DEUS-FUR has allowed us to determine the characteristics of the BAO extrema by directly measuring of the peaks and dips in the BAO spectrum. In this case the choice of a wiggle-free spectrum necessary to define the BAO is of particularly importance. Here, rather than using fitting formulae that may introduce spurious effects and absorb part of the BAO signal, we have computed a smooth spectrum from a N-body simulation with an initial wiggle-free spectrum. 

Such characteristics make the BAO from DEUS-FUR BAO an optimal reference for testing semi-analytical model predictions and calibrate the amplitude of non-linear effects expected in a standard $\Lambda$CDM cosmology. To this purpose we provide in Appendix~\ref{fit} polynomial fitting formulae of the BAO spectrum valid in the redshift interval $0<z<1$, based on a decomposition of the non-linear effects in term of a shift of the BAO extrema, a damping factor and a broadband slope. 

We have tested two widely used models, Halofit \citep{smith03} and RegPT \citep{taruya12}. We find that Halofit overestimates the amplitude of the BAO by a large factor ($1.25$, $2$ and $3$ for the three first peaks respectively). In contrast RegPT accurately reproduce the BAO spectrum up to $k=0.17$~h Mpc$^{-1}$ beyond which discrepancies with respect to DEUS-FUR diverge. This is mostly due to inaccurately predicting the broadband slope of the non-linear power spectrum. Thus, subtracting a smooth spectrum as predicted using RegPT leads to much smaller deviations.

We have detected a small non-trivial coupling between BAO and the broadband slope. Furthermore, we have quantified the shift and damping of each BAO extrema and determined their evolution as function of redshift. We find that at a given redshift the second and third trough are not affected by non-linearities and thus can provide unbiased measurements of the cosmic distances. The redshift dependence of the shift is proportional to $D_+^2(z)$, hence measurements of the shift of the BAO extrema at different redshifts may test the linear growth factor. However, given the relatively small amplitude of the shift ($<1\%$) such measurements may turn to be very challenging even with future surveys. In contrast, the damping of the BAO amplitude has much larger impact and varies with redshift proportionally to $D_+(z)$. Hence, by measuring the ratio of peak-to-trough amplitude at two different redshifts, it should be possible to constrain the growth factor and infer additional cosmological constraints besides those encoded in the cosmic distance to the BAO. 

The non-linear effects on the BAO spectrum vary with the underlying cosmological model. In a future work we will investigate such dependency extending this analysis to the two other DEUS-FUR simulations of non-standard Dark Energy models.

\section*{Acknowledgements}
We are particularly grateful to the team of engineers at TGCC for technical support. We are very thankful to Romain Teyssier (RAMSES code), Simon Prunet \& Christophe Pichon (MPGRAFIC and POWERGRID) for providing their original applications. We are also thankful to Ravi Sheth, Martin Crocce, Francisco Prada, Chia-Hsun Chuang, Atsushi Taruya and Fabrice Roy for fruitful discussions. This work was granted access to HPC resources of TGCC, CCRT and IDRIS through allocations made by GENCI (Grand \'Equipement National de Calcul Intensif) in the framework of the ``Grand Challenges'' DEUS and DEUSS. The research leading to these results has received funding from the European Research Council under the European Community's Seventh Framework Programme (FP7/2007- 2013 Grant Agreement no. 279954). We acknowledge support from the DIM ACAV of the Region Ile de France.

\appendix
\section{Fitting formula for low redshift Baryonic Acoustic Oscillations }
\label{fit}
Here, we provide a polynomial fitting formula to the DEUS-FUR $\Lambda$CDM-W7 BAO spectrum valid in the range $k=0.03$ to $0.3$~h Mpc$^{-1}$ in the redshift interval $z=[0,1]$. Our template is inspired by that used in \citet{seo10} and formulated such as to preserve our decomposition of non-linear effects in terms of shift, damping and continuum. When using functional fitting formula the physical interpretation of this quantities may depend on the exact form of the assumed template, hence the comparison with equivalent quantities derived from other fitting functions should be carefully performed. For this reason, in our analysis of the BAO we have preferred to measure the shift, damping and the continuum directly on the BAO spectrum. Nevertheless a fitting formula is still useful for model testing or comparison with noisy data. An implementation in various languages (fortran 90, C, IDL, etc.) can be found at \url{http://www.deus-consortium.org/overview/results/bao/}.

Our template reads as
\begin{eqnarray}
%\begin{split}
P_{\textrm BAO}^{fit}(k)&=&\left\{P_{\textrm lin}\left[\frac{k}{\alpha(k)}\right]-P_{\textrm smooth}^{\textrm lin}\left[\frac{k}{\alpha(k)}\right]\right\}\times\nonumber \\
 &\times&\exp \left[-\frac{\gamma^2(k) k^2}{2}\right]+\beta(k)
\label{eqfit}
%\end{split}
\end{eqnarray}
with the shift, the continuum and the damping given by
\begin{eqnarray}
\alpha(k)&=&\alpha_0+\alpha_1 k+\alpha_2 k^2\\
\nonumber\\
\beta(k)&=&\beta_0+\beta_1 k+ \beta_2 k^2 + \beta_3 k^3\\
\nonumber\\
\gamma(k)&=&\gamma_0+\gamma_1 k + \gamma_2 k^2
\end{eqnarray}
and the redshift evolution of the coefficient is given as function of the scale factor $a$ in matrix form
\begin{eqnarray}
\left(\begin{array}{c}
\alpha_0\\
\alpha_1\\
\alpha_2\\
\end{array}
\right)&=&\left(
\begin{array}{cccc}
 1.01294   &     -0.0494738   &    0.0336815 \\ 
 -0.250750   &     0.925866   &      -0.562989 \\ 
 0.909234   &      -3.16226   &     1.84910 \\ 
\end{array}
\right) \\
&\cdot&
\left(\begin{array}{c}
1\\
a\\
a^2\\
\end{array}
\right)
\nonumber\\
\left(\begin{array}{c}
\beta_0\\
\beta_1\\
\beta_2\\
\beta_3\\
\end{array}
\right)&=&\left(
\begin{array}{cccc}
 207.558   &       -601.965   &      473.922 \\
  -2985.62  &      8710.23   &       -7156.44 \\ 
 14419.3   &       -42314.9   &      34675.7 \\ 
 -22627.2   &      66883.1   &       -54111.9 \\ 
\end{array}
\right)
\cdot
\left(\begin{array}{c}
1\\
a\\
a^2\\
\end{array}
\right)
\nonumber\\
\left(\begin{array}{c}
\gamma_0\\
\gamma_1\\
\gamma_2\\
\end{array}
\right)&=&
\left(
\begin{array}{cccc}
   -0.478928   &       11.6805   &      -4.05990 \\ 
 7.93381   &      26.9037   &       -14.1445 \\ 
 -7.40005   &       -89.3044   &      48.3370 \\ 
\end{array}
\right)
\cdot
\left(\begin{array}{c}
1\\
a\\
a^2\\
\end{array}
\right)\nonumber
\end{eqnarray}

Statistical errors are given by the number of independent modes in a given $k$-bin as discussed in the text, these are added in quadrature to a systematic error which we have estimated to be $1\%$.

The quality of this fitting formula is shown as orange line in the BAO plots and reproduces well the amplitude, shift and average of the DEUS-FUR BAO extrema.  

For sake of completeness we also provide here the polynomial fitting formula of $P_{\textrm{smooth}}^{\textrm{FUR}}(k)$ at $z=0$: 
\begin{eqnarray}
\textrm{log}[P_{\textrm{smooth}}^{\textrm{FUR}}(k)]&=&2.35243-3.49263\times\textrm{log}(k)\nonumber\\
&-&14.0599\times\textrm{log}(k)^2-43.6840\times\textrm{log}(k)^3\nonumber\\
&-&81.0737\times\textrm{log}(k)^4-96.2789\times\textrm{log}(k)^5\nonumber\\
&-&73.0723\times\textrm{log}(k)^6-33.9677\times\textrm{log}(k)^7\nonumber\\
&-&8.76417\times\textrm{log}(k)^8-0.958067\times\textrm{log}(k)^9\nonumber\\
\end{eqnarray}
We have implemented an extension of this fit to higher redshift assuming the same form as above. A code for this fitting function is also available at:\url{http://www.deus-consortium.org/overview/results/bao/}.

\end{document}